\documentclass[namedreferences]{solarphysics}
\usepackage[hyperref,optionalrh,showbiblabels]{spr-sola-addons} 
\usepackage{graphicx}        
\usepackage{amssymb}        
\usepackage{color}           
\usepackage{breakurl}        

\begin{document}

\begin{article}

\begin{opening}

\title{Automated identification of coronal holes from synoptic EUV maps}

\author[addressref={aff1},corref,email={amr.hamada@mail.com}]{\inits{A.}\fnm{Amr}~\lnm{Hamada}}
\author[addressref={aff1},corref,email={timo.asikainen@oulu.fi}]{\inits{T.}\fnm{Timo}~\lnm{Asikainen}}
\author[addressref={aff1},corref,email={ilpo.virtanen@oulu.fi}]{\inits{I.I.}\fnm{Ilpo}~\lnm{Virtanen}}
\author[addressref={aff1},corref,email={kalevi.mursula@oulu.fi}]{\inits{K.}\fnm{Kalevi}~\lnm{Mursula}}

\address[id=aff1]{ReSoLVE Centre of Excellence, Space Climate Research Unit, University of Oulu, Finland}

\runningauthor{Hamada et al.}
\runningtitle{Coronal hole identification}

%


\begin{abstract}

Coronal holes (CH) are regions of open magnetic field lines in the solar corona and the source of fast solar wind.  
Understanding the evolution of coronal holes is critical for solar magnetism as well as for accurate space weather forecasts. 
We study here the extreme ultraviolet (EUV) synoptic maps at three wavelengths (195{\AA}/193{\AA}, 171{{\AA}} and  304{\AA}) 
measured by {\it Solar and Heliospheric Observatory/Extreme Ultraviolet Imaging Telescope} (SOHO/EIT) and 
 {\it Solar Dynamics Observatory/Atmospheric Imaging Assembly } (SDO/AIA) instruments.
The two datasets are first homogenized by scaling the SDO/AIA data to the SOHO/EIT level by means of histogram equalization.
We then develop a novel automated method to identify CHs from these homogenized maps
by determining the intensity threshold of CH regions separately for each synoptic map.
This is done by identifying the best location and size of an image segment, which optimally
contains portions of coronal holes and the surrounding quiet Sun allowing us to detect the momentary intensity threshold. 
Our method is thus able to adjust itself to the changing scale size of coronal holes and to temporally varying intensities.
To make full use of the information in the three wavelengths we construct, a composite CH distribution,
 which is more robust than distributions based on one wavelength. Using the composite CH dataset we discuss the 
temporal evolution of CHs during the solar cycles 23 and 24.

\end{abstract}

\end{opening}

%
 \section{Introduction}
     \label{S-Introduction} 
Coronal holes (CHs) are regions characterized by lower electron density and temperature than the typical quiet Sun (QS). CHs were first observed by  \cite {Waldmeier_1957}  who noticed long-lived regions of negligible intensity in coronagraphic images of the 5303{{\AA}} green line. 
The earliest EUV observations of coronal holes were made by \cite{Tousey_1968}, who noted from spectroheliograms obtained by rocket experiments that EUV emission in polar regions seemed to be weaker than in the surrounding regions.
In the 1970s, CHs were observed as discrete dark patches in the X-ray and EUV solar disk images \citep{Krieger_etal_73, Zirker_1977}.  
CHs coincide with rapidly expanding open magnetic fields with one predominant magnetic polarity \citep{Wang2004}. They are the source region of high speed solar wind streams (HSS) that play an important role for geomagnetic activity \citep{Krieger_etal_73, Neupert-Pizzo_1974, Tsurutani_etal_1995, Gosling1999, Miralles2010, Holappa_2014, Mursula2015}  and are the most important driver of energetic particle precipitation into the Earth's atmosphere \citep[e.g.,][]{Asikainen2016}. CHs do not erupt suddenly like solar flares but evolve rather slowly, emitting a rather constant and fast solar wind for several solar rotations. Detailed reviews of the plasma and magnetic properties of coronal holes have been given by \cite{Zirker_1977}, \cite{Harvey_Sheeley_1979} and \cite{Cranmer_2009}. 
 
 It is difficult to uniquely and objectively identify coronal holes on solar images. They appear differently in different wavelengths, which makes it
somewhat arbitrary to use one wavelength alone to determine them. The area, shape, and darkness of any given hole are not the same in different observing wavelengths \citep{Kahler_etal_83, Harvey_2002}. The CH boundary is not sharp and is often partly obscured by magnetic structures related to surrounding active regions. Coronal holes, both polar and equatorial, have historically been identified visually and hand-traced by experienced observers. However, different human observers typically determine the CH boundaries differently. Moreover, manual detection methods are extremely time-consuming and problematic in quantitative studies.

Development of automated methods of detecting and identifying solar features has increased dramatically in recent years driven by the need to process a growing volume of data. More recently, there have been several attempts to automate the identification and detection of coronal holes using different techniques, such as edge-based segmentation \citep{Scholl_2008}, perimeter tracing \citep{Kirk_2009}, multichannel fuzzy clustering \citep{Barra2009} and intensity thresholding \citep{Krista_2009, deToma_2011}. Methods based on a fixed intensity threshold of one wavelength \citep{Abramenko2009} have been found to be insufficient \citep{deToma_2011} because the threshold intensity of CHs may vary with time. 
Using observations in the He-I 10830{{\AA}} line which depicts CHs as bright regions, \cite{Henney2007} determined the CH intensity
threshold effectively as the 85th percentile of intensity and classified pixels with intensity above this threshold as CHs.
They also checked the unipolarity of the CH candidates using photospheric magnetograms. 
\cite{DeToma2005} used a fixed threshold method on multiple wavelengths (four SOHO/EIT lines, He-I 10830 {{\AA}}, magnetograms and H-$\alpha$ images) to determine stringent criteria for a CH region. \cite{deToma_2011} used a similar technique for synoptic maps, but with a variable threshold level from one map to another. The threshold was determined from the width of the Gaussian distribution of pixel intensities, with parameters depending on wavelength. \cite{Scholl_2008} used histogram equalization and fixed thresholding to first find the low-intensity regions in the three wavelengths of SOHO/EIT, and then the statistics of magnetic field parameters measured by SOHO/MDI to distinguish between filaments and CHs. Similar methodology was used by \cite{Krista_2009}, however with the difference that they used local histograms of SOHO/EIT, STEREO/EUVI, and Hinode/XRT images to find the low-intensity regions.

Although the CH intensity threshold is fundamentally important when determining CH boundaries, many of the methods discussed above chose the CH threshold rather arbitrarily. Moreover, most studies identifying CHs only consider a relatively short period of data
and use only a single wavelength to identify CHs. In this paper, we present a rigorous automated CH detection method, which uses information
from three different wavelengths, is robust for temporally varying CH threshold intensity (either naturally or due to changes in detector
 properties or intensity scaling of EUV synoptic maps) and automatically takes into account
the changing scale size of coronal holes when determining the threshold intensity. 
We apply the method to the synoptic EUV maps of SOHO and SDO from 1996 until present. The paper is organized as follows. In Section \ref{S-Data} we briefly describe the solar EUV observations by SOHO/EIT and SDO/AIA instruments and the properties of each dataset. 
In Section \ref{S-Threshold} we present the method for the automated CH identification. 
Section \ref{S-CH maps} presents the treatment of CH maps and the way of differentiating coronal holes from filament channels. 
In Section \ref{S-CH evolution} we analyze the CH properties obtained using different wavelengths. 
In Section \ref{S-CHB} we compare the evolution of CHs and photospheric magnetic field.
Section \ref{S-Discu} summarizes the obtained results.

\section{Data}
     \label{S-Data} 

In this work, we use EUV synoptic maps constructed from SOHO/EIT and SDO/AIA observations at three different wavelengths 
(195{{\AA}}/193{{\AA}}, 171{{\AA}} and 304{{\AA}}). A total of 276 Carrington rotations (CR 1911-2186), covering the period from 
1996 June 28 to 2017 February 06 were utilized in this paper. This interval includes the solar minima between cycles 22/23 
and 23/24, full solar cycle 23 and the ascending to early declining phase of cycle 24. 

We also use magnetic field synoptic maps to calculate the magnetic properties of coronal holes and to differentiate them from 
filament channels (FC). Synoptic maps of both SOHO/MDI and SDO/HMI were obtained from the Stanford University 
(http://hmi.stanford.edu/data/synoptic.html). The SOHO/MDI (SDO/HMI)
 maps have a resolution of 3600$\times$1080 (3600$\times$1440, respectively)
 pixels and are expressed in equally spaced longitude and sine latitude. The SOHO/MDI maps also incorporate a polar field interpolation, 
which fills the data gap at the polar region of the less visible solar hemisphere 
(this visibility effect for EUV maps is discussed in more detail later). To compare with the EUV synoptic maps the magnetic synoptic maps were
interpolated to equally spaced solar latitude corresponding to the latitudes of the EUV maps.

\subsection{SOHO/EIT synoptic maps} %

Synoptic maps based on the data from SOHO/EIT \citep{Delaboudiniere1995} give a view of the solar atmosphere in four different wavelengths in 1996-2010. Spectral emission lines Fe IX/Fe X (171{{\AA}}), Fe XII (195{{\AA}}), Fe XV (284{{\AA}}), and He II (304 {{\AA}}) provide temperature diagnostics in the range from $6\times 10^4$ K to $3\times 10^6$   (see Table \ref{T1}). However, the high-temperature Fe XV 284{{\AA}}  images from EIT are problematic to analyze because, at solar minimum, hot plasma is mostly absent, resulting in poorly defined 284{{\AA}} images \citep{Kahler2002}. Thus, only the Fe XI/Fe X 
171{{\AA}} and FE XII 195{{\AA}} lines and the He II 304{{\AA}} line are used here. 
The characteristics of the three selected wavelengths are summarized in Table \ref{T1} \citep[see also][]{Moses1997, Petkaki_2012}. The He II 304{{\AA}}  images are dominated by emissions from structures of the transition region network \citep{Benevolenskaya2001}, indicating the magnetic footpoints of coronal loops and outlining the basis of coronal holes. The Fe IX/Fe X 171{{\AA}} images display background emissions that are present over most of the quiet Sun,
while the most intense emissions come from active regions with closed magnetic field. Especially in this wavelength the coronal holes and filament channels 
often appear almost equally dark. The Fe XII 195{{\AA}} images are also dominated by emission from the closed magnetic field regions of the Sun, showing the inner solar corona with different distributions of intensities for coronal holes and the quiet Sun. \\

SOHO/EIT EUV synoptic maps have been constructed by \cite{Benevolenskaya2001} from CR 1911 (1996 June 28) to CR 2055 (2007 March 31) by concatenating strips that are $16^\circ$ wide in longitude, centered at the central solar meridian. We denote this time range as EIT-1 period. 
They have binned the original $1024\times 1024$ pixel images 
to $512\times 512$ pixel resolution, and each pixel with rectangular coordinates has been transformed to the Carrington coordinate system. The resolution of these maps is $1^\circ\times 1^\circ$ of heliographic latitude and longitude covering all longitudes ($1^\circ$ to $360^\circ$) 
and most latitudes ($-83^\circ$ to $+83^\circ$). 
The latitudes have been restricted to this range because the polar regions are not visible throughout the year due to the 
Earth's annually changing heliographic latitude, also called the vantage point. These EIT/EUV synoptic maps are provided by Stanford Solar Observatories Group. Note that the synoptic maps for CR 1911 (1996 June 28) - CR 2042 (2006 April 10) are based on calibrated data, and for CR 2043 (2006 May 08) - CR 2055 (2007 March 31) on preliminary calibrated data. \\

Another set of EIT/EUV synoptic maps for the same wavelengths, covering CR 2058 (2007 June 21) - CR 2102 (2010 October 03) are provided by Space Weather Lab at George Mason University. We denote this time range as EIT-2 period.
 In these maps, longitudinal strips of $13.63^\circ$, centered on the central meridian, have been concatenated using four images per day so that the strips overlap each other by 3/4 of their width. The pixel intensities were calculated as averages of photon counts in the overlapping pixels \citep{HessWebber2014}. The size of these synoptic maps is $3600\times 1080$ pixels ($0.1^\circ \times 0.1667^\circ$). We note that for some reason quite many synoptic maps are missing from this dataset (23 out of 45). It is unclear why this is the case,
even though EIT has been returning data regularly during this period.

\subsection{SDO/AIA synoptic maps} %

Synoptic maps based on the data from SDO/AIA EUVI sensor were obtained for CR 2097 (2010 May 20) - CR 2186 (2017 January 10) from Space Weather Lab at George Mason University \citep{Karna2014}. This time period will be denoted as the AIA period. These maps have the size of $3600\times 1080$ pixels and cover all longitudes and all measured latitudes at best up to $\pm$ $90^\circ$. SDO/AIA EUV wavelengths are listed in Table \ref{T1}.\\

While SOHO is located at the L1 point, SDO is on a geosynchronous orbit. The SDO/AIA images are therefore less contaminated by galactic cosmic rays or solar proton events (SPE), which can deteriorate image quality. For AIA observations in 193{{\AA}}, the measurements are primarily from the Fe XII ($1.6\times 10^6$ K) emission, with additional emissions from Fe XXIV whose formation temperature is about $10\times 10^6$ K. This difference in the temperature response functions between SOHO/EIT 195{{\AA}} and SDO/AIA 193{{\AA}} is unlikely to lead to large differences in coronal hole detection \citep{Caplan2016}. \\

\subsection{Homogenization of the Space Weather Lab maps}

Inspecting the synoptic maps from the different data sources revealed that the pixel values in the synoptic maps represent
the pixel intensities in very different ways. The pixel values in the maps offered by Stanford Solar Observatories represent the
true pixel intensity in linear scale. However, the synoptic maps of SOHO/EIT and SDO/AIA provided by Space Weather Lab (SWL) at
 George Mason University display considerable variability in 
the units of intensity. In most of these maps the pixel values have been given as integer numbers between 0 and 255.
Inspecting the histogram of these pixel values we came to the conclusion that the values in EIT maps of 195{{\AA}} and 304{{\AA}} were roughly
proportional to logarithmic pixel intensity. On the other hand, for the 171{{\AA}} EIT maps the values are not restricted to between 0 and 255,
and seem to be proportional to the linear pixel intensity as in the Stanford maps. The pixel values in the AIA maps of all three wavelengths 
from the SWL are roughly proportional to logarithmic pixel intensity with the exception of the last 6 Carrington rotations (CR 2181-2186) of the dataset,
which are in linear scale and the values are not restricted to between 0 and 255. 
The SWL maps also fill the pixels that correspond to latitudes that are out of view because of the 
vantage point effect with a value of 232. Because this same value can also appear as a real pixel value in bright active regions,
were set pixels with value 232 only at latitudes poleward of 8.3$^\circ$ to NaN and excluded them from all analyses.

Because of these inhomogeneities we first transformed all linear-scale SWL maps to logarithmic scale and then rescaled the 
logarithmic pixel values of all these maps to the logarithmic scale of the Stanford maps averaged over CRs 1999-2055
(roughly years 2003-2007). During this time period the SOHO/EIT intensity distributions are relatively stable (see Figure \ref{F1}).
The rescaling was computed by transformation of normally distributed log-intensities as
\begin{equation}
I^*=\frac{I-\mu}{\sigma}\sigma_{ref}+\mu_{ref},
\end{equation}
where $I^*$ is the rescaled SWL logarithmic intensity and $I$ is the unscaled SWL pixel value in logarithmic intensity scale,
$\mu$ and $\sigma$ are the mean and standard deviation of the SWL logarithmic pixel values from the time period used for scaling
and $\mu_{ref}$ and $\sigma_{ref}$ are the mean and standard deviation of the logarithmic Stanford maps over CRs 1999-2055.
For the EIT maps of Space Weather Lab $\mu$ and $\sigma$ were computed using all the EIT maps (CRs 2058-2102)
and for the AIA maps we used the three first Carrington rotations in the AIA dataset (CRs 2098, 2101 and 2102).
Note that there is considerable temporal drift in the intensity distribution of AIA maps as shown below, which limits the useful
time period for this comparison.

Due to the different resolution of the EIT and AIA synoptic maps, we scaled 
all synoptic maps provided by Space Weather Lab onto the same $1^\circ\times 1^\circ$ longitude-latitude grid as used by Stanford EIT maps.
This was done by averaging logarithmic pixel values in $10\times 6$ pixel blocks to correspond to $1^\circ\times 1^\circ$ angular resolution. 
In the analysis of all synoptic EUV maps we will use logarithmic pixel values.

\subsection{Properties of homogenized synoptic maps} %

\begin{figure} 
\centerline{\includegraphics[width=1.0\textwidth,clip=]{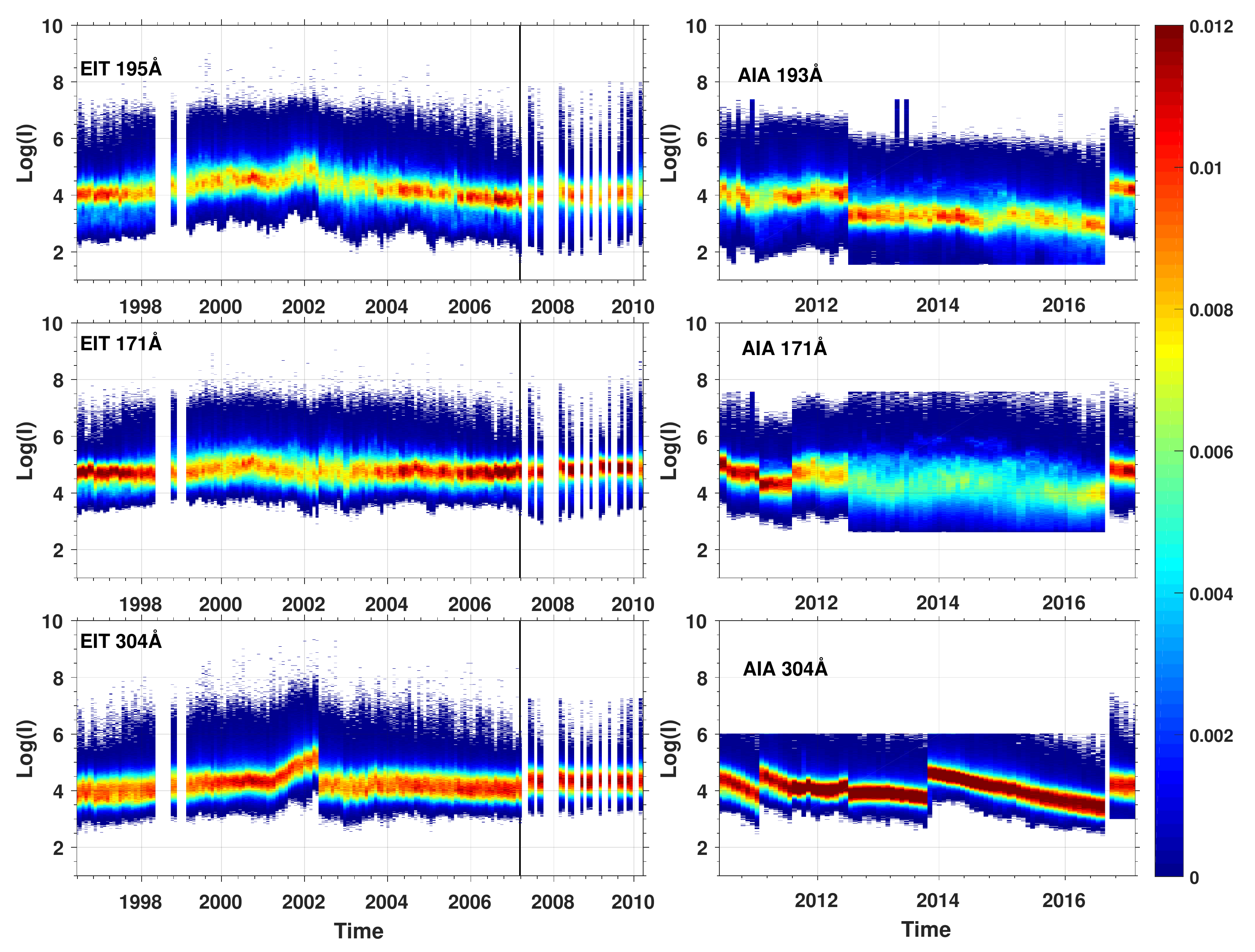}}
\caption{Logarithmic intensity histograms for homogenized SOHO/EIT (left) and SDO/AIA (right) synoptic maps for the three spectral lines 
(195{\AA} /193{\AA}, 171{\AA} and 304{\AA}). All maps are in $1^\circ\times 1^\circ$ resolution. 
Vertical white spaces indicate missing data. Vertical solid line represents the separation between different data sets.}\label{F1}
\end{figure}

We have analyzed all available EUV synoptic maps of EIT and AIA instruments from CR 1911 (1996 June 28) to CR 2102 (2010 October 03) 
and from CR 2097 (2010 May 20) to CR 2186 (2017 January 10), respectively (see Table \ref{T2}). Figure \ref{F1} shows the intensity histograms of SOHO/EIT and SDO/AIA for all synoptic maps separately. A fixed bin width of 0.01 of logarithmic intensity in the histogram was used for all data sets.
The bin values range from 1.5 to 10 in logarithmic intensity. The color indicates the relative number of pixels in each bin so that the
sum of all bins of one CR adds to 1. The maximum of the intensity histogram is typically formed by logarithmic pixel intensities of about 4, 
which is typical for the quiet Sun. AIA 304{{\AA}} histograms for CR 2097 (2010 May 20) to 2179 (2016 July 03) show a constant upper limit which is due to the histogram being cutoff at the maximum pixel value of 255 in the original maps. 
Synoptic maps for the last six AIA maps are also shifted to slightly higher intensity level. 
Intensity histograms for all three AIA wavelengths show an abrupt level change at CR 2125 (2012 June 21). 
Sudden jumps in the intensity distribution are visible in all the wavelengths, e.g., in April 2012.
Some of these jumps are likely related to instrument bakeouts \citep[][Table 1]{Boerner2014}, 
where the instrument is heated in hopes of regaining sensitivity 
lost due to degradation. There are also significant drifts related to fast instrument degradation, which is most clearly
visible in the 304{\AA} wavelength \citep{Boerner2014}.

\begin{figure} 
\centerline{\includegraphics[width=1.0\textwidth,clip=]{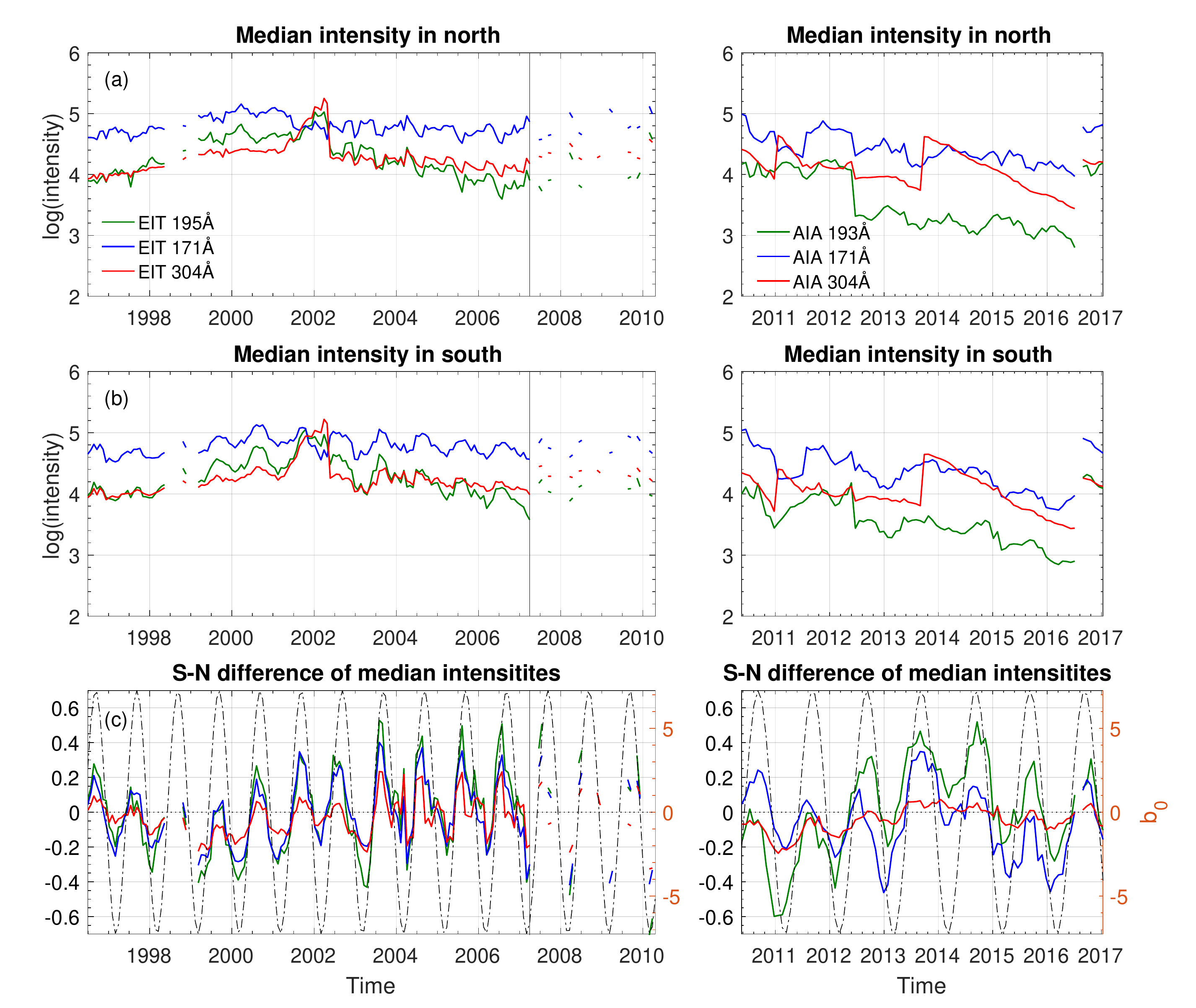}}
\caption{Median logarithmic intensities of SOHO/EIT (left) and SDO/AIA (right) synoptic maps for the three different wavelengths 
separately for the (a) northern, (b) southern hemisphere, and (c) the south-north difference. 
Dash-dotted line indicates the Earth's heliographic latitude angle $b_0$ (right axis). 
Vertical solid line represents the separation between different data sets.}\label{F2}
\end{figure}

Figure \ref{F2} shows the median logarithmic intensities for each CR, separately for the northern and southern hemisphere, 
as well as the south-north difference. The median intensities and particularly their difference show large annual variations, 
roughly in phase with the variation of the Earth's heliographic latitude (so called  $b_0$-angle). The tilt of the solar 
equator by $7.25^\circ$ with respect to the ecliptic plane gives the Earth a better view of the northern (southern) solar 
pole during fall (spring). The $b_0$ angle also affects the observed intensities in the following way: during fall (spring) 
the length of the line-of-sight path through the coronal matter to the southern (northern) hemisphere increases, 
which increases (decreases) the observed intensity of EUV emission. Moreover, active regions surrounding the coronal 
holes obscure the visibility of coronal holes in the less visible hemisphere especially when the line-of-sight angle 
(between local surface normal and Sun-Earth line) is large. 

In addition to the annual variation one can see in the EIT EUV data a variation over the solar cycle 23, with median intensities 
increasing from 1996 to a maximum and then decreasing thereafter. In 2002, a notable increase in the intensities of both 
the hotter corona (195{{\AA}}) and transition region/chromosphere (304{{\AA}}) is observed, while the upper transition 
region/quiet corona intensity (171{{\AA}}) did not show any enhancement at that time, but had its maximum somewhat earlier in 2000. 
This peak in 2002 was first found by \cite{Floyd2003} and later verified in several solar UV/EUV and
other parameters, like F10.7cm radio flux, Mg II K line intensity \citep{Floyd2005} and even in the ionospheric response to solar UV/EUV
\citep{Lukianova2011}.

The median intensities recorded by the AIA instrument (right hand panels in Fig. \ref{F2}) show similar problematic behaviour
related to calibration and degradation issues as the intensity distributions in Figure \ref{F1}.
Therefore, no solar cycle variation of EUV intensities could be found for cycle 24. Note, however, the solar cycle evolution of the S-N
difference with larger intensities in the northern hemisphere before the solar maximum and in the southern hemisphere after
the maximum. This is valid for both solar cycles 23 and 24, and follows the similar well known pattern of sunspot activity \citep[e.g.,][]{Vernova2002}.

\section{Determining the coronal hole intensity threshold}
     \label{S-Threshold} 
		
\begin{figure} 
\centerline{\includegraphics[width=1.0\textwidth,clip=]{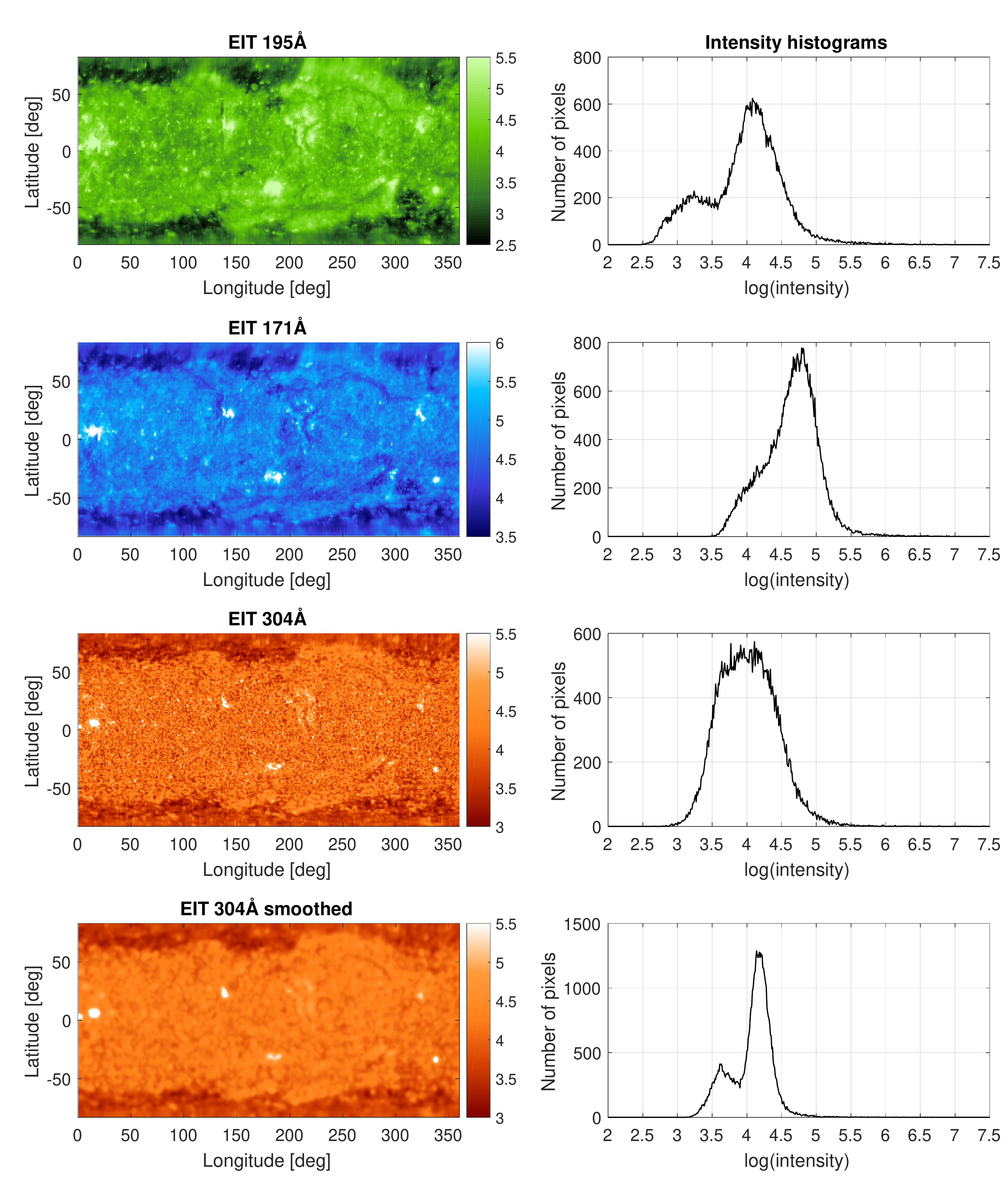}}
\caption{SOHO/EIT synoptic maps (left) and the corresponding intensity distributions (right) for CR 1922 (1997 April 24)
for the three wavelengths (195{{\AA}}, 171{{\AA}} and 304{{\AA}}). The bottom row depicts the spatially smoothed 304{\AA} wavelength map. 
The x-axis of the histogram indicates the intensity values in log-scale and the y-axis shows the corresponding number of pixels.}\label{F3}
\end{figure}
		
The top three rows of Figure \ref{F3} show SOHO/EIT synoptic maps for the three wavelengths (195{{\AA}}, 171{{\AA}} and 304{{\AA}}) 
for CR 1922 (1997 April 24)  and the corresponding histograms of logarithmic intensities. 
The histograms here were computed with the same logarithmic binning as those in Figure \ref{F1}. 
These maps show that CHs can have different contrasts with 
respect to surrounding regions in different wavelengths. 
Even though large polar coronal holes in both hemispheres are evident in all maps, the full image intensity histograms
indicate the coronal holes very differently. 
The green 195{{\AA}} map has the greatest contrast between active and inactive regions leading to a bi-modal intensity distribution,
where the peak at lower intensities corresponds to coronal hole regions. 
The histogram of the blue 171{{\AA}} map looks quite different, depicting a skewed intensity distribution with a single peak. 
The red 304{{\AA}} map has the lowest contrast and an almost unimodal intensity distribution. The smaller structures of the 
chromospheric network seen with the 304{{\AA}} line have very dark and bright features, which produce a similar contrast as coronal holes,
and make the distinction between these features and CHs difficult using this line. 
Therefore, before further processing, the chromospheric 304{{\AA}} maps were first 
spatially smoothed using a 2-D Gaussian filter, with a standard deviation of 2 pixels, thus smoothing out the excessive spatial detail. 
The bottom row of Figure \ref{F3} shows the smoothed EIT 304{\AA} synoptic map and the corresponding intensity histogram.
 One can see that the smoothing reduces the contrast of the small scale features, which are present over most of the solar surface. 
As a result, the large, dark polar CHs become better visible in the intensity histogram as a separate peak.

\begin{figure} 
\centerline{\includegraphics[width=1.0\textwidth,clip=]{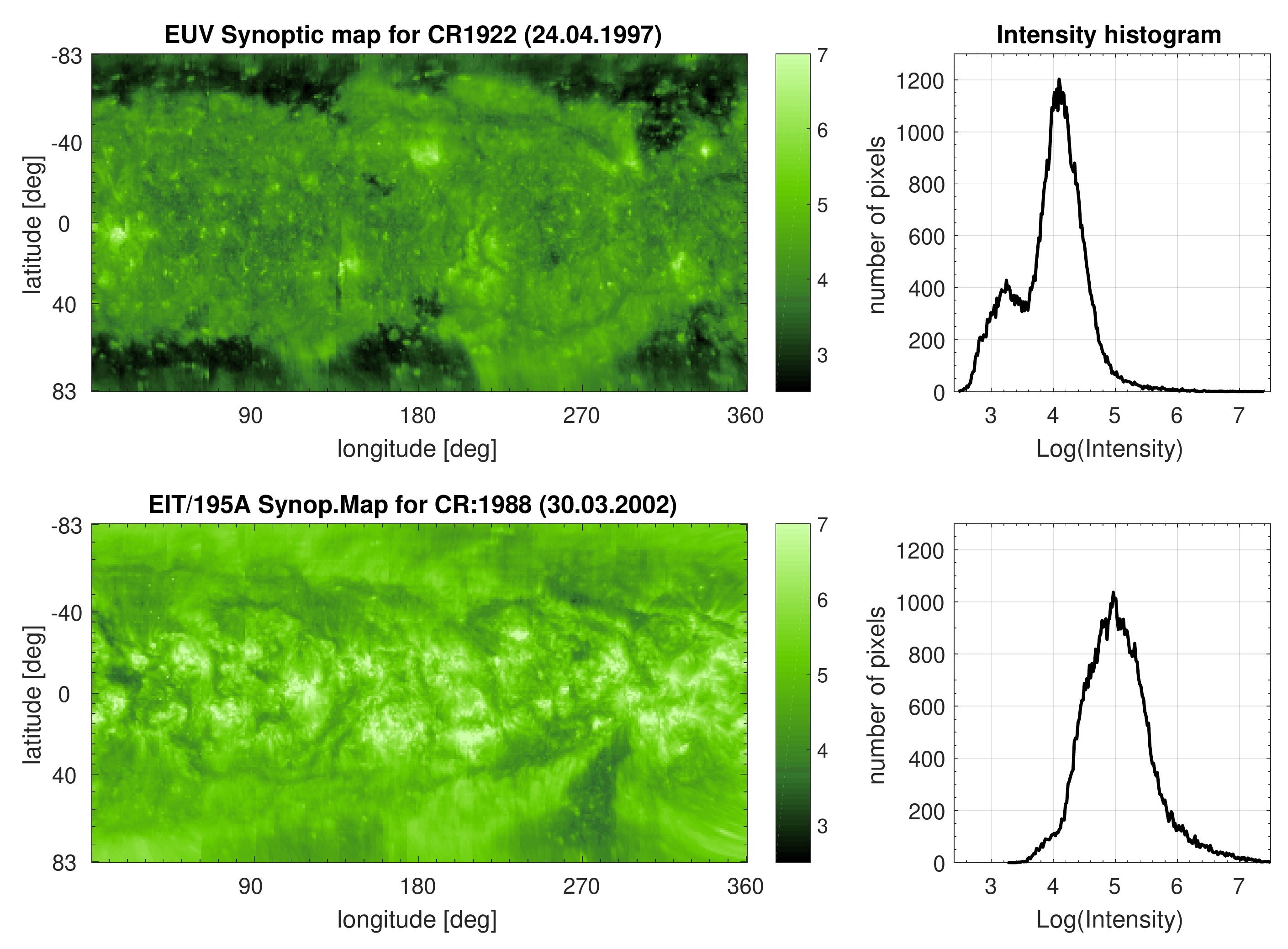}}
\caption{Intensity histogram of SOHO EIT 195{\AA} synoptic map for CR 1922 (1997 April 24) at solar minimum (upper panels) 
and for CR 1988 (2002 March 30) at solar maximum (lower panels).}\label{F4}
\end{figure}

Figure \ref{F4} shows the effect of solar activity on the intensity of the EIT 195{{\AA}} line by comparing the synoptic maps during
 a low activity period in CR 1922 (1997 April 24) and a high activity period in CR 1988 (2002 March 30).  
In general, the range of intensities in EUV maps varies from the high intensity of active regions (ARs) around solar 
maximum (like in 2002) through the intermediate intensity of the quiet Sun (QS) to the low intensity of coronal holes. 
The QS contains typically the largest number of pixels in most images. It is well known that the intensity histogram of 
a solar image often has a bimodal structure where the three major features (CH, QS or AR) may correspond to one mode of the 
intensity histogram \citep{Gonzalez2002, Gallagher1998, Gallagher1999, Belkasim2003}. Especially, when a relatively large part of the image is 
covered by low-intensity regions (CHs), the resulting intensity histogram may show a clear bimodal distribution 
(Figs. \ref{F3} and \ref{F4}, upper panels). As the number of active regions (solar activity) increases, 
the fraction of AR related pixels and the total EUV intensity increase. At the same time, the number and area of 
coronal holes tend to decrease. Then the intensity histogram typically shows a more unimodal distribution 
(Fig. \ref{F4}, lower panels). Comparing the intensity histograms at different solar activity levels, 
it is obvious that the intensity distribution moves to higher intensities with increasing solar activity.  
Also, with increasing solar activity, the CH boundaries get less accurate due to bright coronal loops partly or 
totally obscuring the CH, as discussed above \citep{Krista_2009}.

\begin{figure} 
\centerline{\includegraphics[width=1.0\textwidth,clip=]{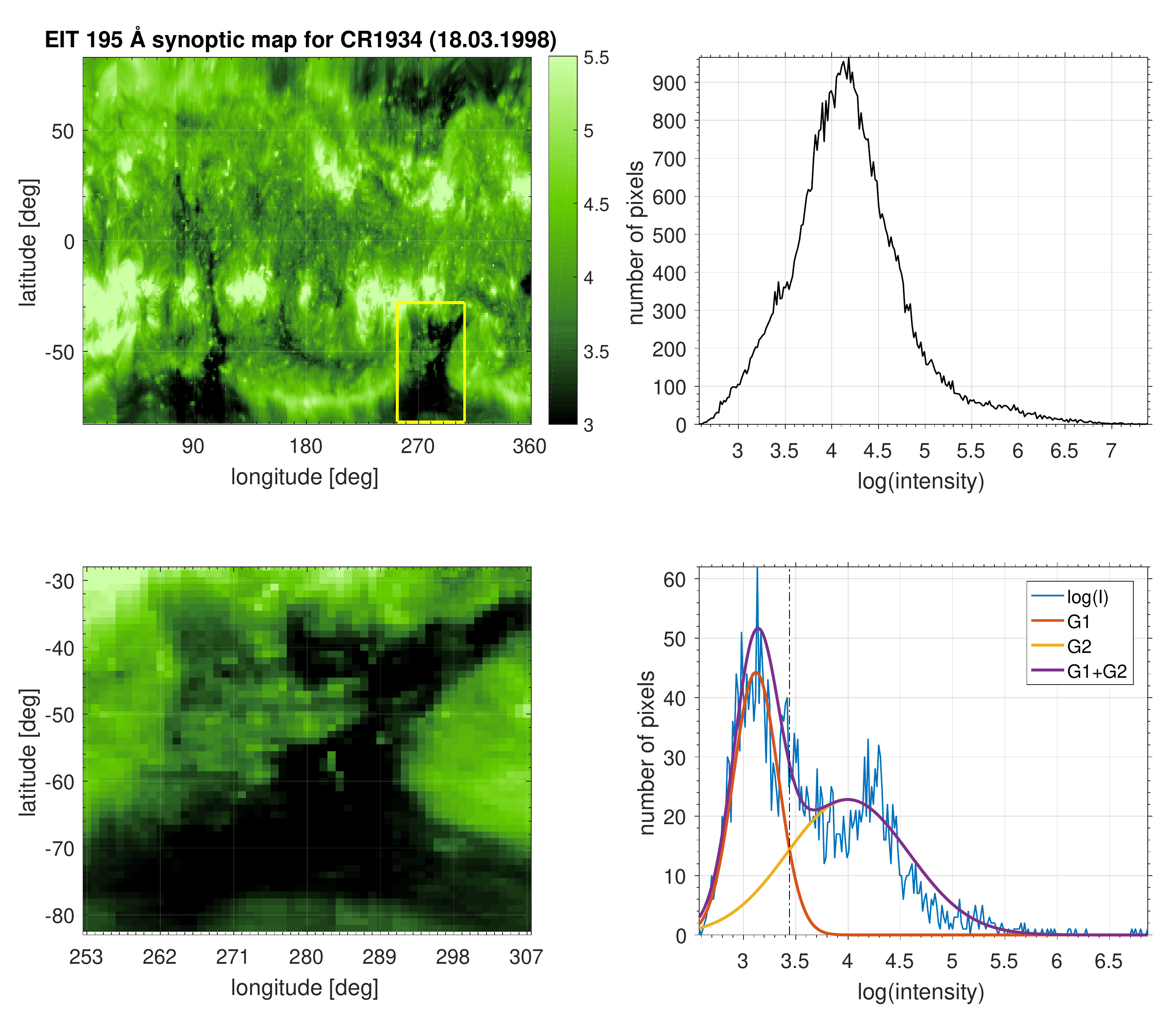}}
\caption{EIT 195 {\AA} synoptic map for CR 1934 (1998 March 18) and its intensity distribution (upper panels). 
Yellow box indicates the optimum segment of length $l=55$; expanded view and the intensity histogram in lower panels. 
Segment histogram was fit to a sum of two Gaussians shown in the lower right panel. 
Vertical dash-dotted line indicates CH threshold intensity.}\label{F5}
\end{figure}

Thus, the histogram of the \textit{entire image} often does not reveal the existence of CHs at times when their total coverage is small. 
For this reason, we aid CH identification by studying the image in small square segments with 
segment length varying from 15 to 55 pixels in 5 pixel steps. 
The identification of CHs is then based on finding such a segment which most clearly depicts the separate 
CH and QS populations in the bi-modal histogram. The best segment for CH identification is the one which has 
as large a contrast as possible and as low a mean intensity as possible. These two criteria aim to ensure 
that there are two separate populations in the histogram and that these two populations represent the
low to intermediate intensities of CHs and QS.  
Figure \ref{F5} depicts an example of an EUV synoptic map (1996 June 29) with an optimal segment denoted by a rectangle with yellow borders. 
The histogram of the segment is clearly bimodal, consisting of two distinct, but partly overlapping 
distributions: the darker CH and the brighter QS. Note also that the histogram of the full map barely shows any clear sign of CHs. 
The contrast of the segment represents the difference between the population of dark pixels and the population of the brighter ones. 
Here, the contrast is defined as RMS contrast, i.e., as the standard deviation of logarithmic pixel intensities:
\begin{equation}
C=\sqrt{\frac{1}{l^2}\sum\limits_{i=1}^l \sum\limits_{j=1}^l \left( I_{ij}-\left<I\right>\right)^2},
\label{eq_C}
\end{equation}
where $I_{ij}$ is the logarithmic intensity at the pixel (i,j) of the square segment of length $l$ and $\left<I\right>$ 
is the corresponding average of log intensities of segment pixels.\\

After many tests, we found the best division of intensities within segments (and the most consistent CH thresholds) 
when using the following score function for a segment:
\begin{equation}
R = \frac{C-\mu_{C}}{\sigma_{C}} - 2 \frac{\left<I\right>-\mu_{I}}{\sigma_{I}},
\label{eq_score}
\end{equation}
where $\mu_{C}$, $\mu_{I}$, $\sigma_{C}$ and $\sigma_{I}$ are the mean values and the standard deviations 
for the contrast $C$ and average logarithmic intensity $\left<I\right>$ of all available segments of given size 
of a particular EUV map, respectively. Each synoptic map was scanned, stepping from pixel to pixel, from bottom-left corner 
to top-right using the square segment. 
In each step, the value of $R$ was calculated. After repeating the calculation for all segment sizes the segment with the maximum 
$R$ value was used to construct the intensity histogram and to determine the CH boundary intensity threshold for the respective rotation (map). 
Note that Equation \ref{eq_C} scales the contrast and average intensity to the same level of standardized units. 
Tests showed that the standardized intensities must be weighted more than standardized contrasts, and that 
the weight of 2 for standardized intensity is an optimum.

\begin{figure} 
\centerline{\includegraphics[width=1.0\textwidth,clip=]{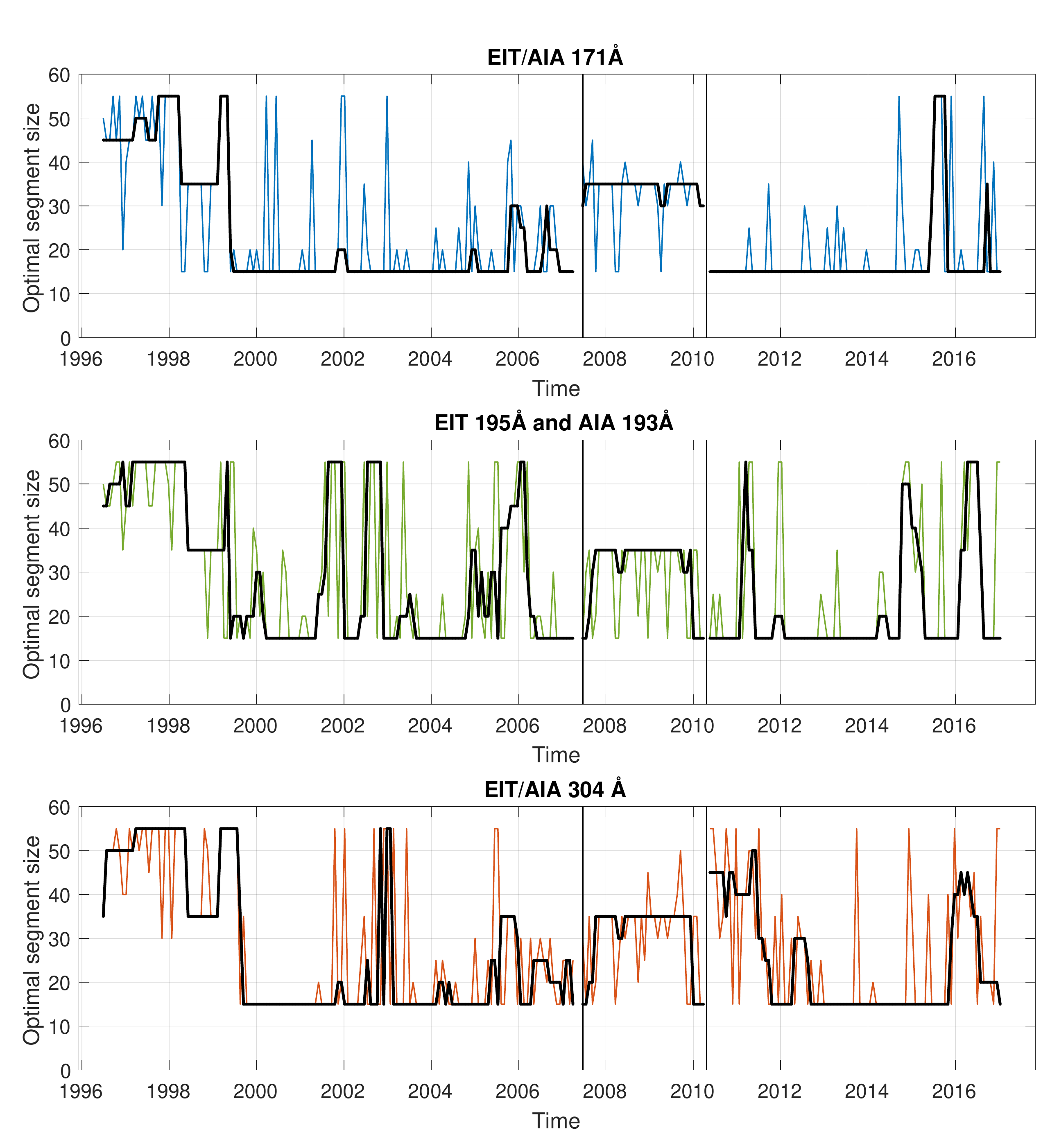}}
\caption{Variation of the optimal segment size with time for the three EIT/AIA wavelengths (171{\AA}, 195/193{\AA} and 304{\AA}).
The black solid lines show the 5-point running median of the optimal segment size.
Vertical solid lines represent the separation between different data sets.}\label{F6}
\end{figure}

Note that finding the optimal segment from segments of different sizes allows the algorithm to adjust to the spatial scale of coronal 
holes, which varies over time. This is demonstrated in Figure \ref{F6}, which shows the variation of the optimal segment size with 
time for the three EIT/AIA wavelengths.
To better see the long-term variation of the segment sizes the plots also show the 5-point running medians (thick black lines).
One can see that the optimal segment sizes generally tend to be larger in solar minima (1996-1998 and 2008-2010) than in solar 
maxima (2000-2001, 2012-2014). Allowing the algorithm to adjust the optimal segment size is important, 
since using a large segment during solar maximum, when CHs are small, would often have the optimal segment of this size to 
contain QS and ARs instead of CHs and QS.

\begin{figure} 
\centerline{\includegraphics[width=1.0\textwidth,clip=]{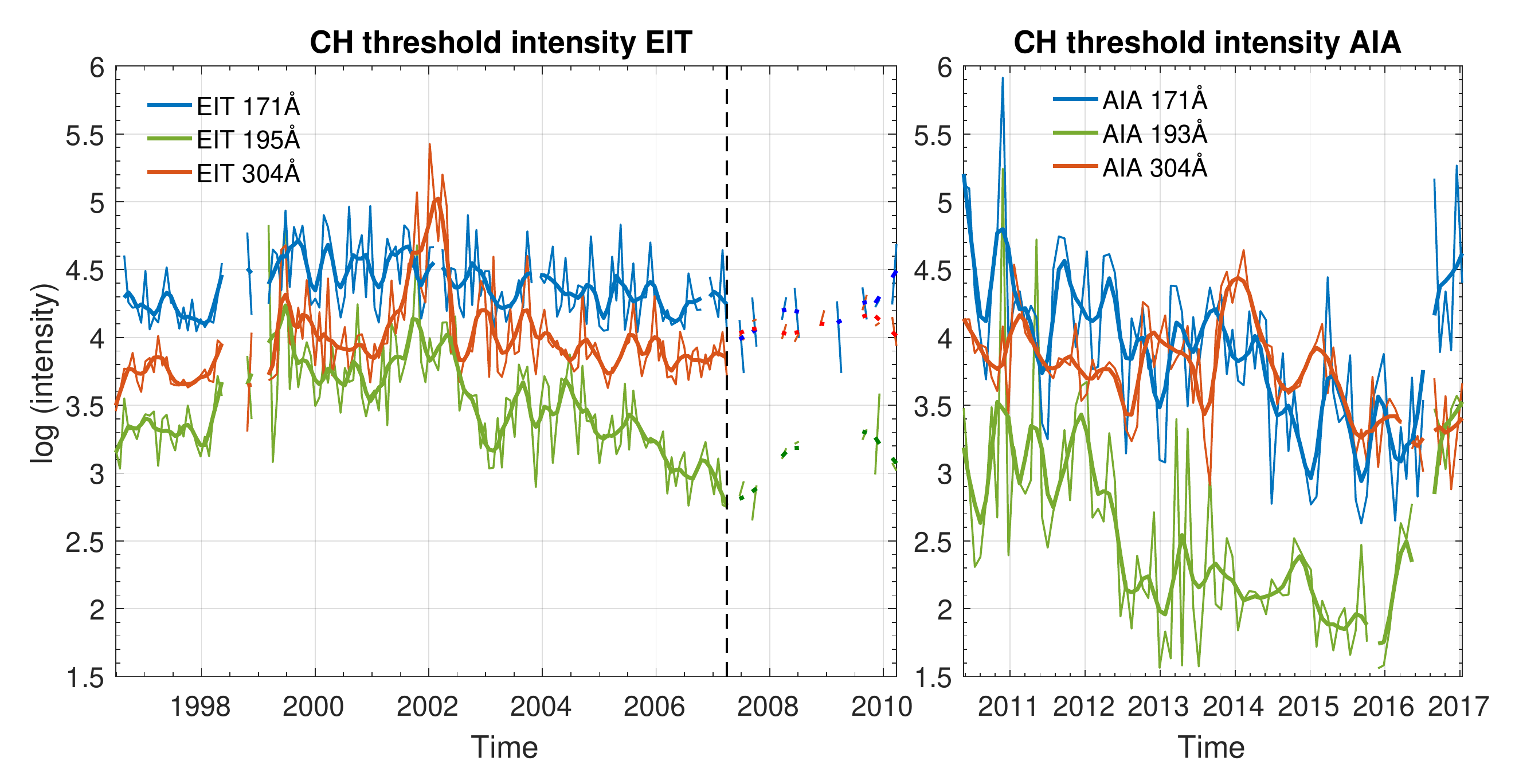}}
\caption{Rotational and 7-CR running mean values of CH threshold intensities for 195{\AA}/193{\AA} (green), 
171{\AA} (blue) and AIA 304{\AA} (red).}\label{F7}
\end{figure}

In order to identify and separate the two populations of the bimodal segment histogram, we fitted the histogram with 
a sum of two Gaussians using the Gaussian mixture model (GMM) implementation of Statistics and Machine Learning Toolbox in MATLAB software. 
The CH threshold intensity was then defined as the intersection between these two Gaussians. 
Even though the Sun is often populated by three separate features (coronal holes, quiet sun and active regions)
fitting a sum of three Gaussians to a relatively small optimal segment would typically result in 
overfitting that does not converge to a meaningful solution. As mentioned above, the method of finding of the optimal 
segment based on Eq. \ref{eq_score} is designed so that it in practice favors those segments, 
which contain roughly equal proportions of CHs and QS and not a third population. 
For a more robust threshold and to reduce the variability from one synoptic map to another, we computed 
the final threshold intensity for a given synoptic map as a running mean of the thresholds of 7 maps centered 
on the CR in question. Figure \ref{F7} shows the CH threshold intensities for the three wavelengths for each CR, 
as well as the final 7-CR running mean values. 
The threshold intensities during the SOHO/EIT Stanford map period vary over the solar cycle 23 roughly similarly 
as the median intensities (see Fig. \ref{F2}). This is expected because the distribution of pixel intensities
of active regions and quiet Sun moves to higher intensity, which in turn moves the intersection of the two Gaussians 
to higher intensities. 
Figure \ref{F7} also shows that the SDO/AIA instrument sensitivity/calibration issues evident in Figs. \ref{F1} and \ref{F2}
are also reflected in the AIA threshold intensities.

 \section{Synoptic coronal hole maps }
     \label{S-CH maps} 
\begin{figure} 
\centerline{\includegraphics[width=1.0\textwidth,clip=]{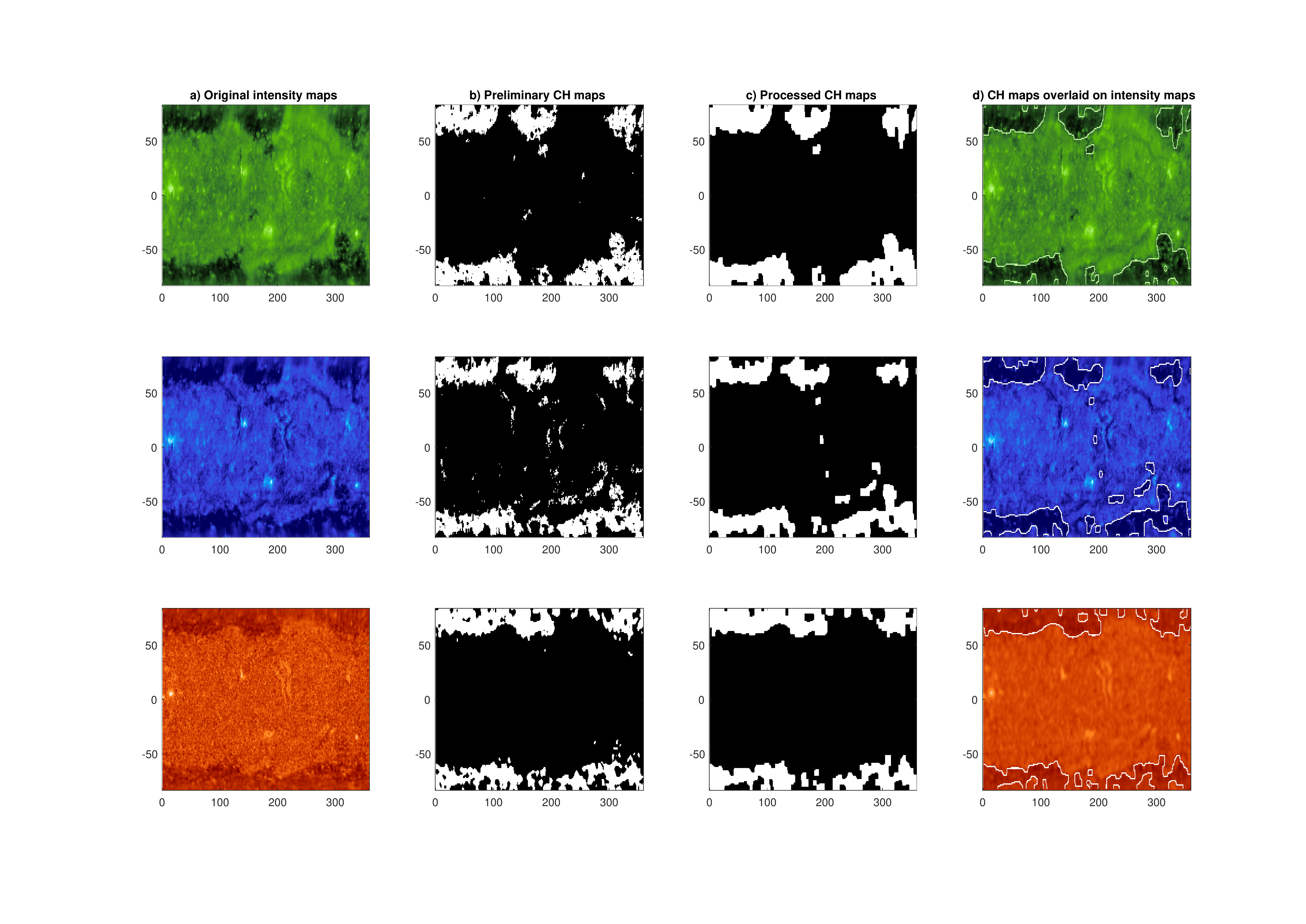}}
\caption{Example of automated CH detection for CR 1922 (1997 April 24) using (a) SOHO/EIT synoptic maps for 195{\AA} (green), 
171{\AA} (blue) and 304{\AA} (red); 
(b) preliminary binary CH maps after determining the CH intensity threshold; 
(c) fully processed CH binary maps; 
(d) edged CH areas overlaid on the EUV maps. The x and y axes depict heliographic longitude and latitude respectively.}\label{F8}
\end{figure}

After determining the final CH threshold intensities, they are used to produce binary maps of CHs by 
marking the pixels with intensity lower than the threshold as 1 and those with higher intensity as 0. 
Examples of such initial binary maps are shown in Figure \ref{F8}b, where white regions indicate coronal holes. 
The binary maps were then post-processed by applying morphological image analysis (MIA) functions of 
dilation (D) and erosion (E) (included in the image processing toolbox of MATLAB). 
Dilation expands each pixel of value 1 to a region defined by a filtering kernel. 
Erosion does the same for pixels with value 0. These operators were applied using a structuring element (kernel) 
that determines the extent of erosion or dilation. The kernel used here is 5$\times$5 pixel square.\\
Using the same kernel, dilation and erosion can be combined to create two other MIA functions open (O) and close (C), which are defined as

\begin{equation}
O(I,K)(x,y)\equiv D[E(I,K),K]
\end{equation}

\begin{equation}
C(I,K)(x,y)\equiv E[D(I,K),K].
\end{equation}

Close operation dilates the image and then erodes the dilated image, while open erodes the image and then dilates the eroded image. 
Close is used to fill the small gaps within the CH areas and to smooth the outer edges of CHs, while open removes 
all the small features having a size less than the kernel's size, without affecting the shape and size of 
larger objects in the image. In this study, we used a filter (F) defined by

\begin{equation}
F(I,K)(x,y)\equiv C[O(I,K),K].
\label{eq_filter}
\end{equation}

Figure \ref{F8}c shows the binary maps after being processed by the filter F. In comparison to the initial binary maps 
shown in Fig \ref{F8}b, most small structures have been removed and the small gaps in CH regions have been filled up.\\

Solar filaments can often be misidentified as coronal holes because they also appear dark on solar EUV images. 
However, filaments can be differentiated from coronal holes by magnetic field. 
Each preliminary CH region obtained as a low intensity region in the EUV map was superimposed with the corresponding 
synoptic map of the radial magnetic field and the relative magnetic flux imbalance was determined for the region. 
Since CHs favour one polarity, while filament channels have a bipolar distribution of magnetic polarities \citep{Scholl_2008}, 
the relative magnetic flux imbalance ($\left<B_r\right> / \left<|B_r|\right>$) of the radial magnetic field $B_r$ within the
initial CH region was used as a parameter to differentiate between CH and FC.
\cite{Scholl_2008} studied CH vs. FC differentiation by testing several criteria: the weighted mean or the sum of field strengths of each polarity, 
the average of the two largest field values of each polarity, the skewness of the magnetic field distribution, 
and the relative magnetic flux imbalance. Among all these criteria, they found that the best discriminator 
was the relative magnetic flux imbalance. Following the study of \cite{Scholl_2008} we labeled the regions 
with less than 15\% imbalance, i.e., with $-0.15 < \left<B_r\right> / \left<|B_r|\right> < 0.15$ as FCs 
and excluded them from the final list of CH regions.  

 \section{Temporal evolution of detected coronal holes}
     \label{S-CH evolution} 
		
\begin{figure} 
\centerline{\includegraphics[width=1.0\textwidth,clip=]{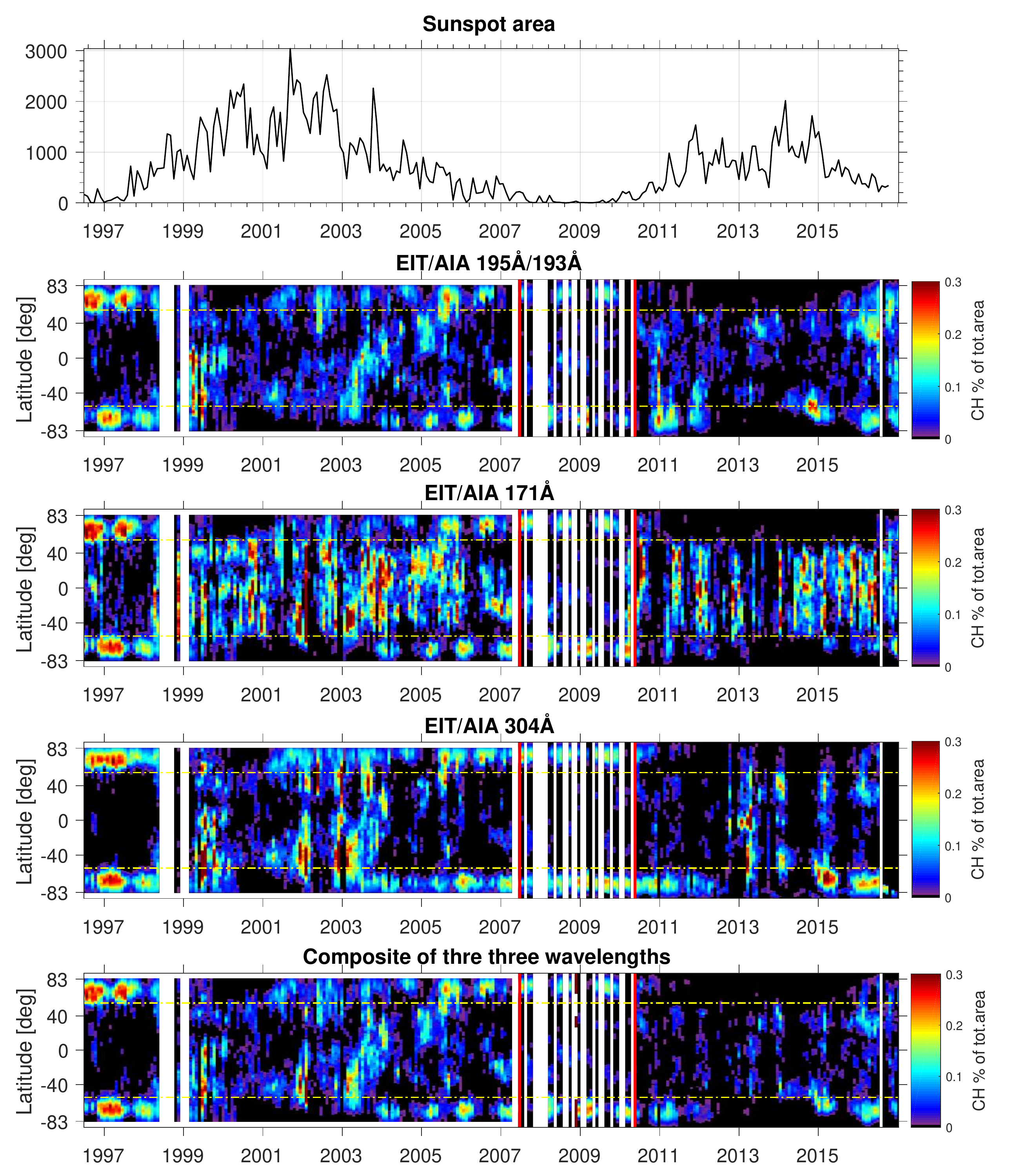}}
\caption{Fractional CH areas at different latitudes in 1996-2017. (a) Monthly sunspot area. (b-d) 
Fractional areas CH for 195{\AA}/193{\AA}, 171{\AA} and 304{\AA}, 
respectively. (e) Composite CH areas computed from the median CH binary maps of the three wavelengths. 
Horizontal yellow dashed lines at $\pm 55^\circ$ latitude distinguish polar and lower-latitude zones.
The red vertical lines show the separation between EIT-1, EIT-2 and AIA periods.}\label{F9}
\end{figure}

We calculated the fractional CH area $f_{CH}(\lambda_i)$ at heliographic latitude $\lambda_i$ relative to the total area of the Sun as 
\begin{equation}
f_{CH}(\lambda_i)=\frac{n_i\cos \lambda_i}{N_i\sum_i \cos \lambda_i}
\label{eq1}
\end{equation}
where $n_i$ is the number of CH pixels at latitude $\lambda_i$ and $N_i$ is the total number of pixels with data 
(i.e., excluding data gaps) at the same latitude. Figure \ref{F9} b)-d) show the latitudinal distribution of fractional 
CH area for the three wavelengths (195{{\AA}}/193{{\AA}}, 171{{\AA}} and 304{{\AA}}). The monthly sunspot areas (obtained from 
https://solarscience.msfc.nasa.gov/greenwch.shtml) are shown 
in Fig. \ref{F9}a) for reference. The depicted time interval from CR 1911 (1996 June 28) to CR 2186 (2017 February 06) 
shows the CH evolution throughout the solar cycle 23 and most of solar cycle 24. 
The horizontal lines at $\pm$ $55^\circ$ latitude are shown here as a division between polar and lower-latitude regions.
The red vertical lines show the separation between EIT-1, EIT-2 and AIA periods. 
Some other definitions for the polar region are also used in literature, with boundary typically defined between $50^\circ$ to $60^\circ$. 
Our definition follows the observation that EUV bright points and coronal green line (Fe XII 5303{{\AA}}) 
emissions are reduced above  $55^\circ$ latitude \citep{McIntosh2014, McIntosh2015}.

During the solar minimum and early ascending phase in 1996-1998 large polar coronal holes (PCH) were detected 
in all three spectral lines. After the solar minimum, PCHs started to diminish and nearly disappeared in both hemispheres in 2000. 
Lower latitude CHs started growing in the ascending phase. Mid- and low-latitude CHs were seen almost throughout 
the whole solar cycle and particularly clearly in the declining phase in 2001-2006. 
The $b_0$ (vantage point) effect is clearly seen in Fig. \ref{F9} in PCH occurrence, 
oppositely phased in the northern (peaks in fall) and southern (peaks in spring) hemisphere. 
This is particularly clear in 1996-1998, but also in 2005-2007. One can also see a large north-south 
asymmetry in PCHs during the solar cycle 24, with the southern PCHs being developed much earlier in 2014-2016 than the northern PCHs.
Similar behaviour was recently observed by \cite{Lowder2017}.

The occurrence of PCHs is fairly consistent in the three different wavelengths, but the distributions of mid- and low-latitude
CHs show larger differences. For example, 171{{\AA}} wavelength indicates much larger low-latitude CHs present in 1999-2007 
and 2010-2017 than the other two wavelengths. The CH areas are somewhat different for the three wavelengths due to different 
scale heights of the emitting plasma. If the emitting plasma has a small scale height, the coronal hole can be more easily discerned.
However, if the scale height is relatively large, the CH boundary is more easily seen at the limb of the Sun, while it is obscured 
at the central meridian \citep{Kirk_2009}. Thus, using the three different spectral lines for CH detection yields a more 
versatile and complementary view of CH areas. 

To make full use of the information about CH regions contained in the three different
wavelengths, and to get a more robust estimate for the CH distribution we computed a composite 
CH distribution from the three wavelengths. The composite was computed by taking the median of the
synoptic binary CH maps of the three wavelengths (first processed with the filter defined in Eq. \ref{eq_filter}).
A pixel in the composite CH map is set as a CH pixel if at least 
two of the wavelengths indicate it as a coronal hole. The latitudinal distribution of the fractional CH areas of composite CH maps is shown 
in Figure \ref{F9}e. One can see that in the polar regions the composite has a closely similar pattern as all three individual wavelengths. 
In the low-latitude region the composite more closely depicts features present in 195/193{{\AA}} and 304{{\AA}} wavelengths,
while the additional CH areas present in 171{{\AA}} are mostly not included in the composite.

\begin{figure} 
\centerline{\includegraphics[width=1.0\textwidth,clip=]{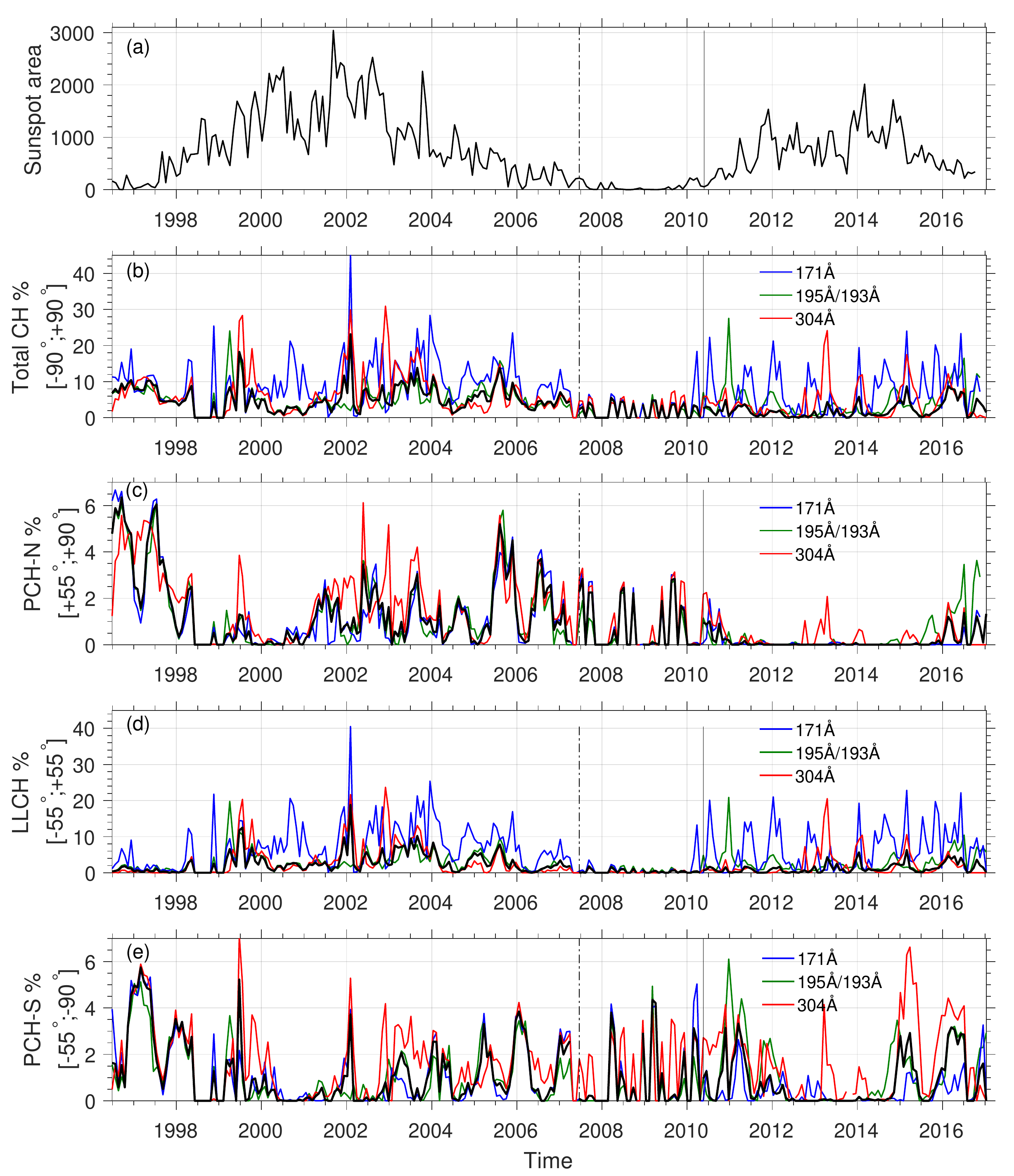}}
\caption{Time series of fractional CH areas for the three spectral lines 195{\AA}/193{\AA} (green), 171{\AA} (blue) 
and 304{\AA} (red). Thick black line corresponds to CH areas calculated from the composite CH maps based on the three
separate wavelengths. (a) Monthly total sunspot area; (b) total CH area in latitude range [+90$^\circ$...-90$^\circ$];
 (c) northern polar CH [+55$^\circ$...+90$^\circ$]; (d) low-latitude CH [-55$^\circ$...+55$^\circ$]; 
(e) southern polar CH  [-90$^\circ$...+55$^\circ$]. The vertical dash-dotted separates EIT-1 and EIT-2 time periods and 
the solid vertical line the EIT-2 and AIA time periods.}\label{F10}
\end{figure}

\begin{figure} 
\centerline{\includegraphics[width=1.0\textwidth,clip=]{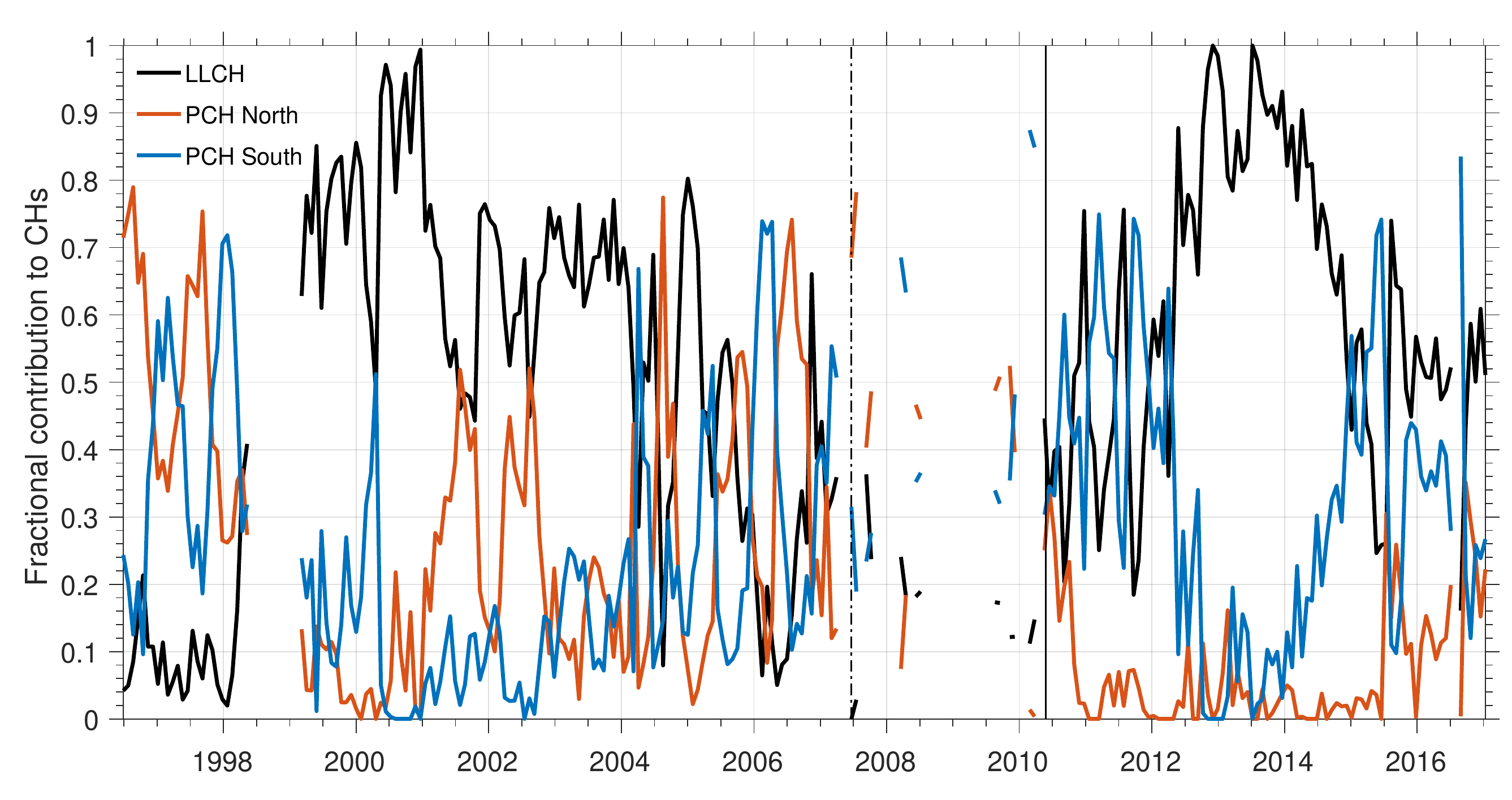}}
\caption{Fractional contribution of mid- to low-latitude CHs (LLCH) and polar CHs (PCH) in the north and south to the total CH area
according to the composite CH distribution. Black line corresponds to LLCH, red line to northern PCH and blue line to southern PCH.}\label{F11}
\end{figure}

Figure \ref{F10}b shows the fractional areas of all coronal holes (total CH between $-90^\circ$...$+90^\circ$) for all three wavelengths
and the composite, while the Figs.
\ref{F10}c-e depict the fractional areas of northern PCHs (above +55$^\circ$), mid- to low-latitude CHs (LLCH, between $-55^\circ$...$+55^\circ$) 
and the southern PCHs (below -55$^\circ$) respectively. 
The fractional area of CHs at any latitude range has been calculated by taking the sum of fractional CH areas (Eq. \ref{eq1}) 
over the respective latitude range. Note the different scale in Figs. \ref{F10}b-e.

One can see that the PCHs in both hemispheres peak around solar minima in 1996-1998, in the declining to minimum phase in 2002-2010 and
in the declining phase in 2015-2017 and have 
a clear minimum around solar maxima in 2000-2001 and 2013-2014. The LLCHs on the other hand peak during the solar maximum and early declining phase
1999-2005,
and have a clear minimum around solar minima. Because PCHs and LLCHs are largely in antiphase but are both quite large in the early declining phase,
the total CH area in Fig. \ref{F10}b shows only a weak solar cycle variation with maximum in 2002-2005. The peak of LLCH area in 2003 (Fig. \ref{F10}d) coincides with the 
largest number of high-speed solar wind streams at 1 AU \citep{Mursula2015, Mursula2017} and energetic particle precipitation at 
that time \citep{Asikainen2016}.

The fractional CH areas depicted in Fig. \ref{F10} and their overall temporal evolution agree fairly well with the recent study by \cite{Lowder2017}, who also used SOHO/EIT and SDO/AIA data to identify coronal holes. 
There is a good agreement between our results and those of \cite{Lowder2017} (their Figure 3) for the polar CH areas.
There is a difference in the relative size of northern PCH areas between 2002 and 2005, where our northern PCH area is clearly larger in 2005, while the two years have roughly equally large northern PCH areas in \cite{Lowder2017}.
Moreover, while we find a small but clear non-zero northern PCH area in 1999, it is practically zero in \cite{Lowder2017}.
However, we find a significant difference for the LLCH areas between the two studies. 
Lowder et al. (2017) indicate that LLCH areas increased rather steadily from 1997 to 2003 without a significant decrease between these years, i.e., around sunspot maximum of cycle 23. 
Contrary to this, our results suggest local maxima in LLCH area in 1999, 2002 and 2003, and a rather long local minimum in 2000-2001. 
There is a notable difference in the size of LLCH area especially in 2000-2001 between these two CH series.

Figure \ref{F11} indicates by what fraction the LLCHs (black) and PCHs in the northern (red) and southern (blue) hemispheres
 contribute to the total CH area at different times. These fractions have been computed from the composite CH distribution.
One can again, even more clearly, see that in solar minima the CH area is dominated by PCHs. In the solar maxima and early declining phase
the CH area is dominated by LLCHs. Figure \ref{F11} also shows clearly the large north-south asymmetry of PCHs 
during the early declining phase of solar cycle 24, which was already evident in the CH distribution of Figure \ref{F9}.

\section{Evaluation of CHs extracted from different wavelengths and datasets}

\begin{figure} 
\centerline{\includegraphics[width=1.0\textwidth,clip=]{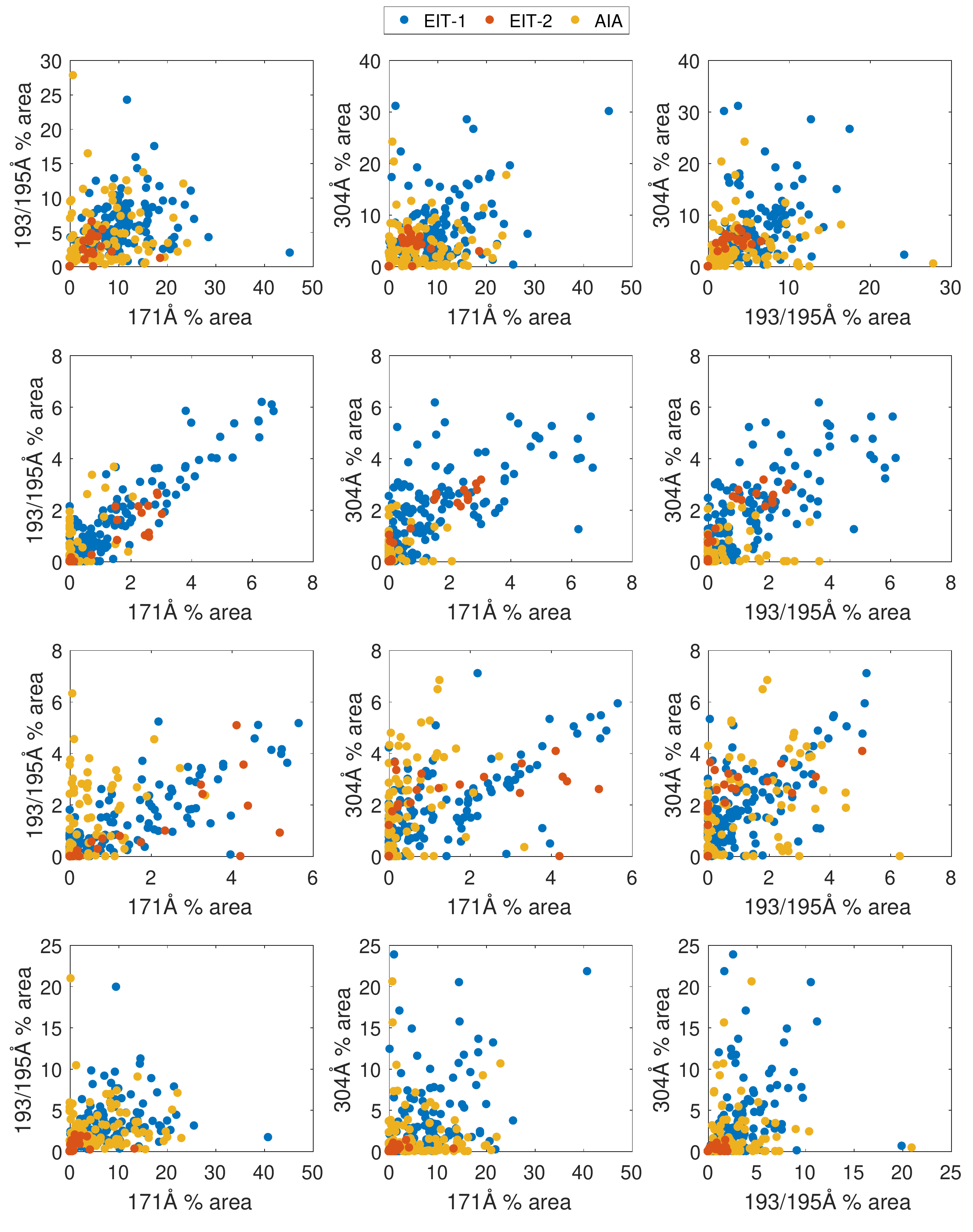}}
\caption{Scatter plots between the fractional total CH areas for the three wavelengths. Top row: total CH fractional area, 2nd row: northern PCH,
3rd row: southern PCH, bottom row: low-mid latitude CH.}\label{F12}
\end{figure}

Figure \ref{F12} shows the mutual scatterplots between the 274 data points (total number of Carrington rotations) of total CH, northern PCH, 
LLCH and southern PCH areas between the three wavelengths in 1996-2017. 
Table \ref{T3} presents the correlation coefficients and statistical significance for each plot. 
For the Stanford synoptic maps (EIT-1 period; blue points in Fig. \ref{F12}) the correlations between 
total CH areas determined from the different wavelengths
are rather modest around 0.32-0.38, but statistically highly significant ($p<10^{-4}$).
These relatively low correlations are largely due to the differences in the low-latitude CHs as seen in Fig. \ref{F12} and Table \ref{T3}.
The PCH areas of EIT-1 period determined from different wavelengths correlate much better (cc between 0.67-0.91)
with the best correlations found between 195{\AA} and 171{\AA}.
Based on this it is clear that the identification of PCHs is much more consistent between the different wavelengths than the LLCHs.

The corresponding correlations for the Space Weather Lab synoptic maps for EIT (EIT-2 period; red points in Fig. \ref{F12}) 
are roughly in line with
those obtained for the EIT-1 period. The LLCH correlations between different wavelengths are again smaller
than the PCH correlations. The correlation coefficients between the total CH areas are higher for the EIT-2 period than for the EIT-1,
but due to the small number of EIT-2 data points this difference is statistically significant only for the correlation 
between 195{\AA} and 304{\AA} wavelengths.
The higher correlation between these wavelengths in EIT-2 period is due to the fact that LLCHs during the EIT-2 period 
(which occurred in late declining phase and solar minimum in 2007-2010)
were very small and the total CH area was dominated by PCHs, which correlate well between different wavelengths.

For the AIA time period (yellow points in Fig. \ref{F12}) the correlations between different wavelengths are generally 
rather poor and in most cases not statistically significant (see Table \ref{T3}).
Significant correlations between all three wavelengths are obtained for the southern hemisphere PCHs only, 
but even there the correlations remain in the range 0.39-0.48 ($p<10^{-3}$). 
The poor correlations in general are partly due to the problems related to AIA instrument degradation and drift with time 
as well as the problems in the Space Weather Lab AIA synoptic maps as discussed above. 
Another reason for the poor correlations in total CH area is the fact that AIA period mostly covers the ascending phase and solar maximum 
of cycle 24, when the CHs were dominated by LLCHs, for which the correlations are insignificant overall. 
Furthermore, the northern PCHs are largely absent
or very small, which also contributes to their poor and statistically insignificant correlation.

We have also quantified the correspondence of CHs determined from each wavelength with the composite CH distribution.
Table \ref{T4} displays for each wavelength the percentage of pixels in the CH synoptic maps (filtered with Eq. \ref{eq_filter}),
 which are not detected as CHs, when the composite CH maps indicate these pixels as CHs. One can see that for the EIT-1 period the percentages for
all the wavelengths are 24-27\% indicating that they are 75\% consistent with the composite CHs.
The EIT-2 period has more variable percentages, with the 171{\AA} wavelength being most consistent (about 94\% of the time) with the composite CHs.
AIA period has considerably worse percentages indicating, e.g., that as much as 77\% of 171{\AA} pixels disagree with the composite.
The 193{\AA} and 304{\AA} percentages are slightly lower (47\% and 51\% respectively), but still indicate a large inconsistency with the composite.
Overall the percentages in Table \ref{T4} indicate that the CH synoptic maps based on EIT in different wavelengths are rather consistent
with each other. Contrary to this, the AIA CH synoptic maps in different wavelengths are much more inconsistent because of the
reasons discussed above.

\section{Comparing CH and magnetic field evolution}
\label{S-CHB}

\begin{figure} 
\centerline{\includegraphics[width=1.0\textwidth,clip=]{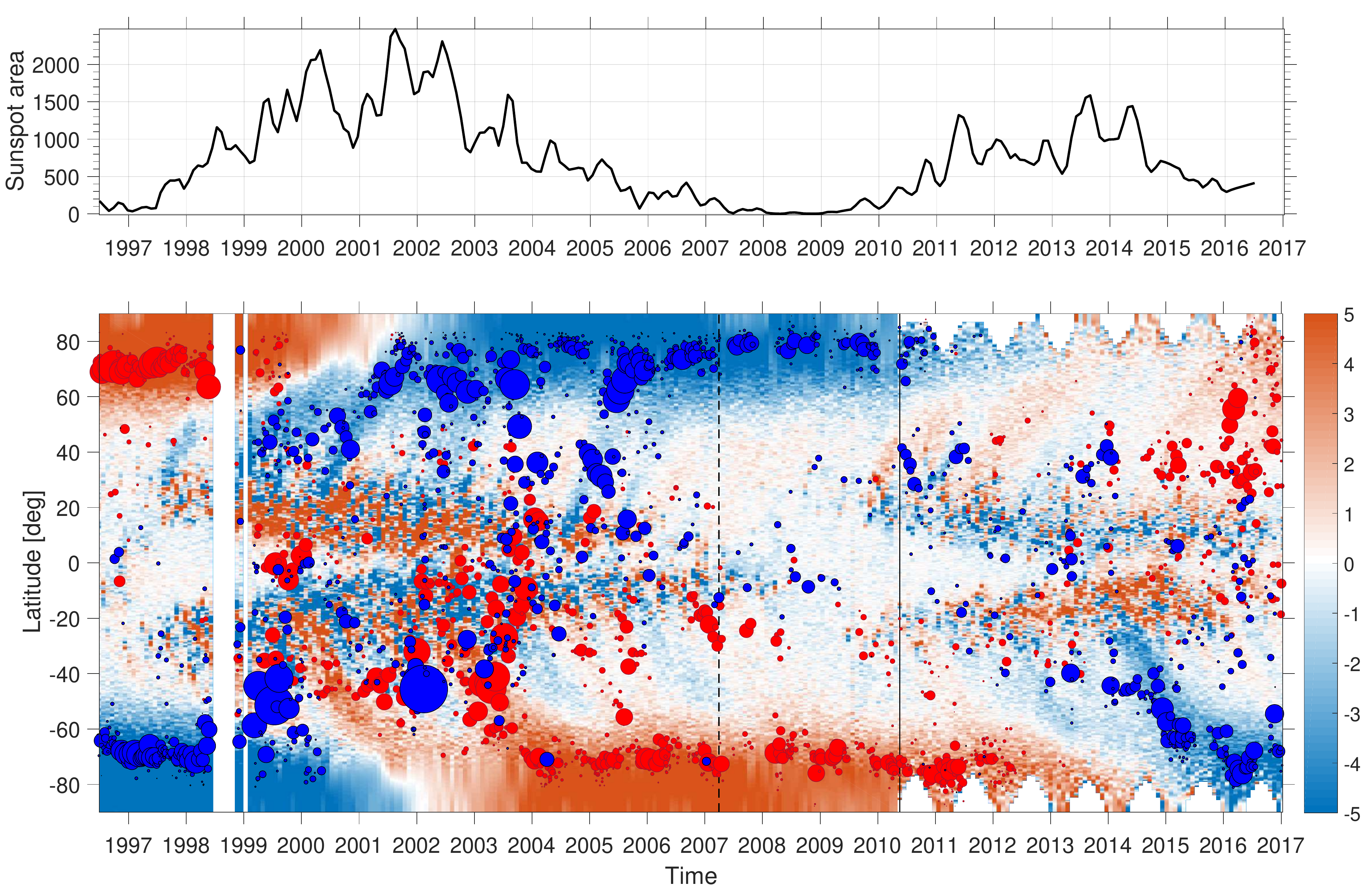}}
\caption{CH evolution through solar cycles 23 and 24. Top panel: total sunspot area for reference. Bottom: The circles depict CH centroids
with size proportional to CH size on the solar surface. The color of the circle indicates
the dominant polarity of the CH (red is positive polarity and blue negative) based on the relative magnetic flux imbalance.
Background color indicates the average photospheric magnetic field intensity in units of gauss. 
The dashed and solid horizontal lines indicate the start of EIT-2 and AIA time periods.}\label{F13}
\end{figure}

Figure \ref{F13} displays the locations of CH centroids (circles) based on the composite CH maps against the measured photospheric magnetic 
field. The size of the circles is proportional to the CH size on the solar surface.
The color of the circle indicates the dominant polarity of the CH (red is positive polarity and blue negative) based on the 
relative magnetic flux imbalance.
The photospheric magnetic field map shows the longitudinal mean of the corresponding (SOHO/MDI or SDO/HMI) synoptic maps. 
The HMI synoptic maps include only the measured regions and no polar field filling is used, while the polar fields in the MDI maps 
have been filled \citep{Sun2015}. The polar field filling method used in MDI effectively removes erroneous pixels and the annual oscillation 
due to the vantage point effect \citep{Virtanen2016}.

The pattern of the CH distribution follows rather closely the evolution of the regions of unipolar magnetic field in the photosphere.
The minimum phase of each solar cycle is characterized by large unipolar fields of opposite polarity at the two poles. 
E.g during the minimum between cycles 22/23 (1996-1999), the northern PCH is dominated by positive magnetic polarity, 
and the southern PCH by negative polarity. Since CHs map the footpoints of open solar magnetic field lines and the 
solar polar magnetic fields are confined to relatively small caps \citep{Sheeley_etal_89}, the polar magnetic field behaves 
in much the same way as the PCHs \citep{Kirk_2009}. However, not all magnetically open field lines must be necessarily 
rooted within the observed CHs \citep{deToma_2011}. 

In the ascending phase of the solar cycle, the CHs along with unipolar regions
of photospheric magnetic flux begin to rush to the poles from mid- and low latitudes, reversing the dominant polarity at high latitudes. 
The formation of new PCHs after the solar maximum results from the transport of new magnetic flux to the solar poles by 
solar meridional circulation and diffusion \citep{Wang2002}. 

The new PCHs with reversed magnetic polarity form at different times in the two hemispheres, 
first the negative polarity PCHs in the north in mid-2001
and then the positive polarity PCHs in the south in 2003. 
The northern PCH reaches its minimum size in June 2012, almost a year 
before the southern PCH in May 2013. The positive polarity magnetic field in the south persists about 2 years longer than the 
opposite polarity field in the north. After the maximum of solar cycle 24 (2014), positive and negative polarity 
PCHs again start to increase. The CH centroids poleward of 55$^\circ$ in Figure \ref{F13} 
indicate that the southern, negative polarity PCH begins to form first (around 2014), almost two years before the northern, 
positive polarity PCH (in 2016). The formation of PCHs during cycle 24 seems to proceed much faster than in the previous 
solar cycle 23. This agrees with observations of \cite{Hathaway2014}, who showed that solar meridional circulation during 
cycle 23 was significantly slower than during cycle 24 and contributed to the weak polar magnetic fields in solar cycle minimum 
at 2010 and to the weak solar cycle 24.  
     
  \section{Discussion and summary}
     \label{S-Discu}      
		
In this paper, we have presented a new method of automatically identifying coronal holes from synoptic EUV maps based 
on SOHO/EIT and SDO/AIA data at three different wavelengths.
The synoptic maps used here were obtained from two 
different data sources, Stanford Solar Observatories Group and Space Weather Lab of George Mason University. 
They span from 1996 to the beginning of 2017, covering the solar cycle 23 and majority of cycle 24. 
The new method introduced here uses these maps to find the optimal threshold intensity for coronal holes for each synoptic map (solar rotation) separately
by determining the best location and size for a segment of the synoptic map that optimally contains portions of 
coronal holes and the surrounding quiet Sun. This allows us to dynamically detect the intensity threshold of each synoptic map. 
The method is able to adjust itself to the changing scale size of coronal holes and to temporally varying 
(either naturally or due to changes in detector properties or intensity scaling of EUV synoptic maps) intensities.
The new method also differentiates the filament channels from coronal holes by the relative magnetic flux imbalance
method introduced earlier \citep{Scholl_2008}. 
We used SOHO/MDI and SDO/HMI synoptic magnetic maps to estimate the magnetic properties of CH identified regions.

The new method was applied to the synoptic EUV maps of the three different wavelengths, which are known to display the CHs somewhat differently. 
Most consistent CH regions between the three different wavelengths were obtained for polar regions, while the low-latitude CH regions were considerably
different. To make full use of the information contained in the different wavelengths we produced
a composite distribution of CHs by taking the median of the morphologically filtered binary CH maps from the three wavelengths. 
This approach yields a more robust result than using measurements only from one wavelength, which most earlier CH detection methods do.

We found considerable differences in SOHO/EIT and SDO/AIA synoptic maps, which partly result from the degradation of the AIA instrument with time.
These problems are already evident, e.g., in the AIA image histograms (Figure \ref{F1}), as jumps and drifts. 
However, the problems related to the AIA EUV maps are largely alleviated by 
taking a robust composite of the three wavelengths. This, e.g., reduces the anomalously high detection rate of low-latitude CHs 
in the 171{\AA} wavelength, which was related to the darkening of these source images with time, 
especially after mid-2012, due to instrument degradation issues and 
consequent clipping of pixel values due to incorrect usage of color table values instead of real pixel intensities (see the Note added below).

The distribution of coronal holes constructed for time period 1996-2017 shows the well known solar cycle variation of coronal holes.
The polar coronal holes peak in the declining phase to minimum of the cycle, while the low-latitude coronal holes peak in solar maximum to 
early declining phase. We observed a significant time lag between the two solar hemispheres in the formation of new polar coronal holes
after solar maxima. In both cycles 23 and 24 the polar coronal holes formed first in that hemisphere which was to have a negative magnetic polarity,
while the polar coronal hole with a new positive polarity field developed only 1-2 years later. We found that the formation of polar coronal 
holes was significantly faster during cycle 24 than during cycle 23, which is consistent with earlier observations of solar meridional 
circulation being slower during cycle 23. We also observed a significant north-south
asymmetry of polar coronal holes during the declining phase of cycle 24, with the southern polar coronal holes being much larger than the northern ones.

The new CH detection method was designed to be as robust as possible against temporal changes in EUV intensity, either natural or artificial, 
and to make full use of the information contained in observations made at different wavelengths. To our knowledge this is the first time that
coronal hole identification is based on such a dynamical and adaptive method, as well as on a combination of 
 different wavelengths into a composite that is more robust than a result based on any single wavelength.

\section*{Note added}
After completing the analysis discussed in this paper it came to our knowledge that the Space Weather Lab synoptic maps, whose pixel values appeared to represent logarithmic pixel intensity in the numerical range from 0 to 255, were actually using 8-bit color table pixel values of the colormap used to plot the images (N. Karna, personal communication, 2017). 
The erroneous synoptic maps have been in public server for several years, and have been used in a few publications, but have recently (November 2017) been corrected.
Fortunately, the colormap used was monochromatic so that the color table values are proportional to the
real pixel intensity. 
Thus, this is not a problem for the novel CH detection method presented in this paper, which relies on contrasts between pixel values.
However, the color table values clip the real pixel values below (above) the lower (upper) threshold, which is clearly seen in the SDO/AIA histograms of Figure \ref{F1}.

\begin{acks}
We acknowledge the financial support by the Academy of Finland to the ReSoLVE Centre of Excellence (projects 272157, 307411) as well as to project 257403. The EUV/magnetogram synoptic map data were obtained from the Stanford Solar Observatories Group (http://sun.stanford.edu/synop/EIT/index.html) and Space Weather Lab at George Mason University (http://space weather.gmu.edu/projects/synop). The monthly sunspot areas were obtained
from the Royal Observatory of Greenwich - USAF/NOAA Sunspot Data center (https://solarscience.msfc.nasa.gov/greenwch.shtml).
\end{acks}
         
%

%
\begin{table}
\caption{Selected SOHO/EIT and SDO/AIA wavelengths}
\label{T1}
\begin{tabular}{cccc}     
\hline               
  Source & Wavelength & Ion & Peak Temperature (MK)\\
\hline               
 SOHO/EIT & 195{{\AA}}  & Fe XII & ~1.6 \\
 SDO/AIA & 193{{\AA}}  & Fe XII & ~1.6 \\
 SOHO/EIT and SDO/AIA & 171{{\AA}}  & Fe IX-X & ~1.3 \\
 SOHO/EIT and SDO/AIA & 304{{\AA}}  & He II & ~0.08 \\               
\hline               
\end{tabular}
\end{table}

\begin{table}
\caption{Data coverage of synoptic maps available for the instruments used. The dates refers to the starting time of observations}
\label{T2}
\begin{tabular}{cccccc}     
\hline               
  Source & Observable & \multicolumn{2}{c}{Start} & \multicolumn{2}{c}{End} \\
              &                    &   CR & Date & CR & Date\\
\hline               
 SOHO/EIT & 195{{\AA}}, 171{{\AA}} and 304{{\AA}}  & 1911 & ~1996.06.28 & 2102 &  ~2010.10.03 \\
 SOHO/MDI & Radial magnetic field                    & 1911 & ~1996.06.28 & 2102 &  ~2010.10.03 \\
 SDO/AIA & 193{{\AA}}, 171{{\AA}} and 304{{\AA}}    & 2097 & ~2010.05.20 & 2186 &  ~2017.01.10 \\
 SDO/HMI & Radial magnetic field                      & 2097 & ~2010.05.20 & 2186 &  ~2017.01.10 \\          
\hline               
\end{tabular}
\end{table}

\begin{table}
\caption{Correlation coefficients between fractional CH areas determined from different wavelengths
separately for the EIT-1, EIT-2 and AIA data periods. All correlation are statistically significant
(with a confidence level of exceeding 99\%), except the ones depicted with {\it italic} font.
The * (+) symbol indicates that the difference between the corresponding values EIT-1 and EIT-2 (AIA) periods
is statistically significant at 5\% level.}
\label{T3}
\begin{tabular}{cccc}     
\hline               
  EIT-1 & 195{\AA} vs. 171{\AA} & 304{\AA} vs. 171{\AA} & 304{\AA} vs. 195{\AA} \\
\hline               
 Total area 	   & 0.32+  &  0.37+ &   0.38*+\\
 North PCH  	   & 0.91+  &  0.67*+&   0.73*+\\
 South PCH  		 & 0.82+  &  0.72+ &   0.70+\\
 Low-latitude CH & 0.37+  &  0.43+ &   0.42+\\
\hline               
 EIT-2 & 195{\AA} vs. 171{\AA} & 304{\AA} vs. 171{\AA} & 304{\AA} vs. 195{\AA} \\
\hline
 Total area 		 & 0.53+  &  0.58+ &   0.85*+ \\
 North PCH  		 & 0.92+  &  0.96+ &   0.9+ \\
 South PCH  		 & 0.76+  &  0.59 &   0.65 \\
 Low-latitude CH & {\it 0.22}  &  0.36+ &   0.53+ \\
\hline               
 AIA & 195{\AA} vs. 171{\AA} & 304{\AA} vs. 171{\AA} & 304{\AA} vs. 195{\AA} \\
\hline
 Total area 		 & {\it 0.08} &  {\it 0.01}&   {\it 0.04} \\
 North PCH  		 & 0.58 &  {\it 0.18} &   {\it 0.19} \\
 South PCH  		 & 0.36 &  0.36       &   0.37 \\
 Low-latitude CH & {\it 0.00} &  {\it -0.05} &   {\it -0.02} \\
\hline
\end{tabular}
\end{table}

\begin{table}
\caption{Percentage of pixels in CH synoptic maps of different wavelength, which do not agree with composite CH map.}
\label{T4}
\begin{tabular}{cccc}     
\hline    
       &  171{\AA} & 193/195{\AA} & 304{\AA} \\           
EIT1	 &  26\%   &   27\% &   24\% \\
EIT2   &  6\%    &   36\% &   15\% \\
AIA    &  81\%   &   50\% &   56\% \\
\hline
\end{tabular}
\end{table}


%

%
%

%
%
 \bibliographystyle{spr-mp-sola}

%
%
%
%

\end{article} 
\end{document}